\documentclass[12pt]{article}
\pdfoutput=1
\usepackage{graphicx,psfrag,epsf,color}
\usepackage{amsmath,amssymb,amsfonts}
\usepackage{array}
\usepackage{cite}
\usepackage{rotating}
\usepackage{slashed, cancel}
\usepackage{xcolor}
\usepackage[caption=false]{subfig}
\usepackage{graphicx}
\usepackage{mathrsfs}  
\usepackage{tabularx}
\usepackage{multirow} 
\usepackage{array}
\usepackage{floatrow}
\usepackage{physics}
\usepackage{color,xcolor}
\usepackage{lscape}
\usepackage[colorlinks,citecolor=blue,urlcolor=blue,linkcolor=blue]{hyperref}

\newcolumntype{C}{>{\centering\arraybackslash}X}
\numberwithin{equation}{section}
\begin{document}
\allowdisplaybreaks

\begin{titlepage}

\begin{flushright}
{\small
CERN-TH-2026-111\\
SI-HEP-2025-20\\
P3H-26-040\\
TTP26-017\\
May 20, 2026
}
\end{flushright}

\vskip1cm
\begin{center}
{\Large \bf\boldmath $B_c \to \eta_c$ form factors at large recoil: \\[0.4em] 
SCET analysis and a three-loop consistency check
}
\end{center}

\vspace{0.5cm}
\begin{center}
Guido~Bell$^a$, Philipp B\"oer$^b$, Thorsten Feldmann$^a$, Dennis Horstmann$^{a,c}$ \\ and Vladyslav Shtabovenko$^a$ \\[6mm]

{\it $^a$Theoretische Physik 1, Center for Particle Physics Siegen, \\
Universit\"at Siegen, 57068 Siegen, Germany}\\[0.3cm]

{\it $^b$CERN, Theoretical Physics Department, CH-1211 Geneva 23, Switzerland}\\[0.3cm]

{\it $^c$Institute for Theoretical Particle Physics, KIT, \\ Wolfgang-Gaede-Straße 1, 76131 Karlsruhe, Germany}
\end{center}

\vspace{0.6cm}
\begin{abstract}
\vskip0.2cm\noindent

The double-logarithmic series of non-relativistic \mbox{$B_c \to \eta_c$} form factors at large recoil is governed by a coupled set of integral equations, reflecting an intricate interplay between arbitrarily many soft-quark and soft-gluon exchanges. Whereas we previously derived these integral equations with diagrammatic resummation techniques, we analyze the form factors in the limit $m_b \gg m_c \gg \Lambda_{\rm QCD}$ with methods from Soft-Collinear Effective Theory (SCET) in this work. Although the resulting factorization theorem for the so-called soft-overlap contribution is known to be spoilt by endpoint divergences, it can still be used at the level of bare (regularized) quantities at any fixed order in perturbation theory. By calculating the required ingredients, we show that the SCET analysis confirms the predictions of the integral equations up to three-loop order. We also argue that the iterative structure and the intertwined soft-quark and soft-gluon effects can be derived from standard renormalization-group equations of the $B_c$-meson light-cone distribution amplitudes, provided their inverse moments are regularized with an appropriate cutoff.
\end{abstract}

\end{titlepage}

\section{Introduction}
\label{sec:Intro}

Precision studies of exclusive $B$-meson decays offer important insights into the quark flavour sector of the Standard Model. The development of QCD factorization (QCDf)~\cite{Beneke:1999br,Beneke:2001ev} at the beginning of the millennium enabled, for the first time, a systematic and controlled approach to compute the hadronic matrix elements for decays into final states containing light (energetic) mesons. From the phenomenological perspective, it turns out, however, that power corrections of $\mathcal{O}(\Lambda_{\rm QCD}/m_B)$ cannot be neglected at the current level of  precision. Despite significant progress, the existing QCDf techniques -- and their reformulation within  the framework of Soft-Collinear Effective Theory (SCET)~\cite{Bauer:2000yr,Bauer:2001yt,Bauer:2002nz,Beneke:2002ph,Beneke:2002ni} -- are not yet capable of addressing these power-suppressed contributions. Technically, the problem manifests in endpoint divergences arising in the convolution integrals between short-distance coefficient functions and the relevant hadronic matrix elements, which are expressed through light-cone distributions amplitudes (LCDA).

Specifically, the hadronic matrix elements for semi-leptonic $B$ decays into a light pseudo\-scalar meson ($P=\pi, K , \eta, \ldots$)   can generally be parametrized by three independent form factors. In the limit of large recoil energy, $E \sim \mathcal{O}(m_B)$, these are related at leading power in the heavy-quark expansion up to calculable factorizable corrections~\cite{Charles:1998dr,Beneke:2000wa}. Schematically, for a given form factor $F_i(q^2)$ with $q^2 = (p-p')^2 \ll m_B^2$, one has  
\begin{align}
\label{eq:BtoPfactorization}
 \langle P(p')| (\bar{q} \, \Gamma_i b)(0)|B(p)\rangle  \equiv F_i(q^2) &= C_i(q^2) \, F(q^2) + \frac{\alpha_s}{2\pi} \, \Delta F_i(q^2) + \mathcal{O}\Big(\frac{\Lambda_{\rm QCD}} {m_B}\Big) \,.
\end{align}
Here, for a given decay mode, $F(q^2)$ is a universal function, often referred to as the \emph{soft-overlap form factor}, and the perturbatively calculable coefficient functions $C_i(q^2)$ include the short-distance QCD corrections from hard loop momenta, $|\ell^\mu|\sim m_b$. The additive terms $\Delta F_i(q^2)$ arise from  gluon exchange between the active quarks in the weak transition and the spectator quark. While the latter can be factorized into short-distance scattering kernels and hadronic LCDA, the hard-scattering picture for the soft-overlap form factor fails, because the arising convolution integrals with the LCDA turn out to diverge at the endpoints where the spectator-quark momentum becomes small.

To shed light onto the physical origin and mathematical structure of these endpoint-divergent convolutions, we  studied the soft-overlap form factor in a perturbative setup in~\cite{Boer:2023tcs,Bell:2024bxg}. More precisely, for heavy-to-light $B_c \to \eta_c$ transitions with the scale hierarchy \mbox{$m_b \gg m_c \gg \Lambda_{\rm QCD}$}, the mesons can be approximated as non-relativistic bound states~\cite{Bell:2005gw}. The quark mass $m_c$ then provides a physical infrared cutoff, and relativistic corrections to the hadronic matrix elements become calculable in perturbation theory. The present article is a sequel to~\cite{Bell:2024bxg}, in which we derived the all-order structure of large double-logarithmic corrections, associated in part with the endpoint divergences, to the soft-overlap form factor. 

The purpose of the present paper is twofold. First, we draw a connection between the diagrammatic analysis of large double logarithms in~\cite{Bell:2024bxg} and an effective-theory description, in which endpoint-divergent moments of LCDA appear in a bare (unrenormalized) factorization formula. Second, we provide an independent consistency check of our previous results up to next-to-next-to-next-to-leading order (N$^3$LO) in perturbation theory, corresponding to the three-loop level. Given the highly intricate structure of the underlying integral equations that govern the double-logarithmic series to all orders~\cite{Bell:2024bxg}, verifying the results at this non-trivial order, using alternative theoretical methods, provides a strong validation for both approaches. 

The remainder of this article is organized as follows. In Sec.~\ref{sec:recap} we briefly review the double-logarithmic structure of the soft-overlap form factor for $B_c \to \eta_c$ transitions at large recoil, which constituted the main result of our previous work~\cite{Bell:2024bxg}. In Sec.~\ref{sec:factorization} we derive the bare factorization formula for the soft-overlap form factor in terms of LO matching coefficients of SCET-2 operators and two- and three-particle LCDA for the $B_c$ and $\eta_c$ mesons. After introducing suitable regulators for the endpoint-divergent convolution integrals, we demonstrate in Sec.~\ref{sec:polecancellation} that the factorization formula correctly reproduces the double logarithms at one-loop order by calculating hard, hard-collinear, soft and collinear corrections from the standard SCET momentum regions. Combining the bare factorization formula with pole-cancellation arguments in a method-of-regions analysis, we furthermore show that there are only two unknown ingredients to reconstruct the double logarithms at two-loop and three-loop order. We choose these as the coefficients of the $1/\varepsilon^4$ ($1/\varepsilon^6$) pole in the purely hard-collinear region at two loops (three loops), which we  compute explicitly using highly automated multi-loop techniques in Sec.~\ref{sec:FOcalculation}. The result of this intricate calculation is finally found to confirm the prediction of the integral equations from~\cite{Bell:2024bxg}, which demonstrates that the bare SCET factorization theorem correctly captures the dynamics relevant for the double logarithms. In Sec.~\ref{sec:discussion} this conclusion is further corroborated by showing that the double logarithmic series can also be derived from standard Lange–Neubert-type evolution equations of the $B_c$-meson LCDA, provided that their inverse moments are regularized by an appropriate cutoff. We conclude in Sec.~\ref{sec:conclusion}, while additional technical details of our analysis are collected in the appendix.

\section{Summary of previous results}
\label{sec:recap}

The soft-overlap form factor for $B_c \to \eta_c$ transitions at large recoil can be defined as
\begin{align}
	\label{eq:soft-overlap}
	F(\gamma) \equiv \frac{1}{2E_\eta} \bra{\eta_c(p_{\eta})} \big( \bar{c} \, \Gamma \, b\big)(0) \ket{B_c(p_{B})} \,, \qquad \text{with} \qquad \Gamma = \frac{\slashed{\bar{n}} \slashed{n}}{4} \,. 
\end{align}
Here, $n^\mu$ is a light-like vector, $n^2 = 0$, that points into the direction of the four-momentum of the energetic $\eta_c$ meson.  Similarly, $\bar{n}^\mu$ is a light-like vector that points into the opposite direction, with $\bar{n}^2 = 0$ and $n \cdot \bar{n} = 2$. Whereas the form factors are often considered as functions of the momentum transfer $q^2 \approx m_B (m_B - 2 E_\eta)$, we use the large boost factor $\gamma = E_\eta/m_\eta = \mathcal{O}(m_b/m_c) \gg 1$ instead. The perturbative expansion of $F(\gamma)$ then develops large double-logarithmic corrections $\sim\alpha_s^{n} \ln^{2n} (2\gamma) \equiv \alpha_s^{n} L^{2n}$ that have been analysed in~\cite{Bell:2024bxg}. To all orders in perturbation theory, it was found that the double-logarithmic series can be written as
\begin{equation}
\label{eq:Ffinaldecomp}
F(\gamma)  \simeq \,
	\xi_0 \,\exp\bigg\{ -\frac{\alpha_s C_F}{4\pi} L^2\bigg\} 
    \left(2 \, \frac{1+\bar{u}_0}{\bar{u}_0^3}\,
f_2(m_2,m_2) + \frac{C_A}{2C_F \bar{u}_0^3} \, \big(1- f_1(m_2,m_2)\big)  -\frac{1}{\bar{u}_0^2}\right) \,, 
\end{equation}
where $\xi_0$ is given in \eqref{eq:xiLO} below, and $u_0 = m_1/m_{\eta}$ (and $\bar{u}_0 = 1-u_0 = m_2/m_{\eta}$) is the mass ratio of the active quark produced at the weak vertex (the spectator quark) and the mass of the $\eta_c$ meson. Note that for the physical $\eta_c$ meson, we have $m_1 = m_2 = m_c$, i.e. $u_0 = \bar{u}_0 = 1/2$. 
Distinguishing between the two quark masses turns out to be convenient, in particular for the consistency checks at higher loop orders, as more independent coefficients can be varied in the numerical evaluation of the two- and three-loop integrals. The functions $f_{1,2}$ in~\eqref{eq:Ffinaldecomp} are auxiliary functions which obey the implicit integral equations
\begin{align}
  f_1(q_+,q_-) &= 1 + \frac{\alpha_s}{4\pi} \int\limits_{q_-}^{p_{2-}} \frac{dk_-}{k_-} \! \int\limits_{m_{2}^2/k_-}^{q_+} \!\frac{dk_+}{k_+} \;  
  e ^{-s(q_+/k_+,p_{\eta -}/k_-)} \,
  \Big[2 C_F  \, f_1(k_+,k_-) \Big] \,,
    \cr
f_2(q_+,q_-) &= 1 + \frac{\alpha_s}{4\pi} \int\limits_{q_-}^{p_{2-}} \frac{dk_-}{k_-} \! \int\limits_{m_{2}^2/k_-}^{q_+} \!\frac{dk_+}{k_+} \;  
  e ^{-s(q_+/k_+,p_{\eta -}/k_-)} 
  \cr
  & \hspace{20mm} \times 
  \bigg[2 C_F  \, f_2(k_+,k_-) 
  + \Big( C_F - \frac{C_A}{2} \Big) f_1 (k_+,k_-)  + \frac{C_A}{2} \bigg] \,,
\label{eq:finalintegralequations}
\end{align}
where $p_{2-}=2\bar{u}_0\gamma m_\eta$ and $p_{\eta-}=2\gamma m_\eta$. These equations represent the main result of~\cite{Bell:2024bxg}, and they reflect a complicated interplay of two distinct sources of large double-logarithmic corrections: Exponentiated soft-gluon effects described by standard Sudakov factors,
\begin{equation}
    s(r_+,r_-) = \frac{\alpha_s C_F}{2\pi} \, \ln r_+ \, \ln r_- \,, 
\end{equation}
and rapidity-ordered soft-quark configurations leading to the recursive structure of the integral equations, which have recently been reanalyzed in the context of energetic muon-electron backward scattering~\cite{Bell:2022ott,Delto:2025epy} (for earlier work see~\cite{Gorshkov:1966qd}).
Whereas the exponentiated double logarithm in the prefactor in~\eqref{eq:Ffinaldecomp} describes the universal Sudakov factor from soft gluons dressing the weak vertex, the exponentials in the integrand arise from additional cusp-like structures within Feynman diagrams with soft-quark configurations.

Although a closed-form solution to these equations is currently unknown, they can easily be iterated to very high orders with the recursion relations for the series coefficients presented in~\cite{Bell:2024bxg}. Writing $F(\gamma) = \sum_{n=0}^\infty \left(\frac{\alpha_s}{4\pi}\right)^n F^{(n)}(\gamma) + \mathcal{O}(m_{\eta}/m_b)$, the first four terms of the double-logarithmic series read 
\begin{align}
\label{eq:xiLO}
	F^{(0)}(\gamma) = \xi_0 \,\frac{2+\bar{u}_0}{\bar{u}_0^3} \,, \qquad \text{with} \qquad \xi_0 = \frac{\alpha_s C_F}{4\pi} \frac{\pi^2 f_{B} f_{\eta} m_B}{N_c E_{\eta}^2 m_{\eta}} \,, 
\end{align}
and 
\begin{align}
\label{eq:FO}
	F^{(1)}(\gamma)  &\simeq \xi_0 L^2 \left( C_F \frac{1+2\bar{u}_0}{\bar{u}_0^3} - \frac{C_A}{2\bar{u}_0^3}\right) \,, \nonumber \\ 
	F^{(2)}(\gamma)  &\simeq 
	\xi_0 L^4 \left( 
    -C_F^2 \,\frac{7+10\bar{u}_0}{6\bar{u}_0^3}  
     + C_F C_A\, \frac{3-2\bar{u}_0}{12\bar{u}_0^3} \right) \,, \nonumber \\ 
    F^{(3)}(\gamma) &\simeq \xi_0 L^6 \left(C_F^3 \, \frac{29+44\bar{u}_0}{90\bar{u}_0^3}
    - C_F^2 C_A \, \frac{1-28\bar{u}_0}{180\bar{u}_0^3}\right) \,,
\end{align}
which reflects a non-trivial dependence on both  the color factors and the quark mass ratio $\bar{u}_0$. In addition, we derived the asymptotic behavior of the soft-overlap form factor in the large-energy limit $E_\eta \to \infty$ in~\cite{Bell:2024bxg}. It turns out that the functions $f_{1,2}$ grow linearly in $\alpha_s L^2$ in this limit, which slightly weakens the overall Sudakov suppression from soft gluons dressing the weak-interaction vertex.

\section{Bare factorization formula in SCET}
\label{sec:factorization}

Heavy-to-light form factors at large recoil are known to receive quantum corrections associated with four different physical modes: Two perturbative modes, \textit{hard} ($\mu_h^2 \sim m_b^2$) and \textit{hard-collinear} ($\mu_{hc}^2 \sim m_b \Lambda_{\rm QCD}$), and two non-perturbative modes, \textit{soft} and \textit{collinear} ($\mu_s^2 \sim \mu_c^2 \sim \Lambda_{\rm QCD}^2$). The derivation of QCD factorization theorems is based on successively integrating out the two short-distance scales in a two-step matching procedure (see e.g. \cite{Bauer:2002aj,Beneke:2003pa,Lange:2003pk}), which we sketch in the following for the soft-overlap form factor. It is well known that this will result in ill-defined, i.e.\ endpoint-divergent, convolution integrals of the coefficient functions and the appearing hadronic matrix elements.  Therefore, the factorization formula presented at the end of this section should be considered in terms of bare quantities, supplemented with suitable regulators to render the integrals mathematically well-defined. While such a bare factorization formula prevents the use of renormalisation-group (RG) techniques to resum the logarithmically-enhanced corrections to all orders, we stress that it can be used to reproduce the QCD result at any fixed order in perturbation theory. We also note that the analysis in this section is independent of the specific details of the non-relativistic setup, and we will therefore use the standard notation employed in the SCET literature, denoting e.g.~the argument of the form factor by $q^2$ instead of $\gamma$, throughout this section.

\subsection{Matching corrections}
\label{subsec:matching}

In a first matching step, one integrates out the hard scale, $m_b \to \infty$, to match onto an intermediate effective theory called SCET-1. In general, $B \to P$ form factors receive contributions from two, so-called $A$-type and $B$-type, operators, see e.g.\ \cite{Beneke:2015wfa}.
The $A$-type current, which defines the soft form factor, can be chosen as
\begin{align}
\label{eq:JAJB}
    J_A = \bar{\chi}_{hc} h_v \,, 
\end{align}
Here $\chi_{hc} = W^\dagger_{hc} \xi_{hc}$ is the gauge-invariant building block for a hard-collinear quark, with $\xi_{hc}$ the large quark-spinor component in the high-energy limit,
and $W_{hc}$ a hard-collinear Wilson line. Similarly, $h_v$ denotes the large component of the heavy-quark field with four-velocity $v$. 
Hadronic matrix elements of the $B$-type current, which contains an additional hard-collinear gluon field,
lead to endpoint-finite convolutions in SCET-2, and they define the factorizable terms $\Delta F_i$ in~\eqref{eq:BtoPfactorization}. 
Contrarily, the inverse moments of LCDA that appear after matching the $A$-type current to SCET-2 are divergent for small light-cone momenta as we will see below. The Dirac projector $\Gamma$ in~\eqref{eq:soft-overlap} is chosen precisely in such a way that only the $A$-type current $J_A$ contributes to the EFT description of the form factor $F$. In the heavy-quark limit, we can thus write
\begin{equation}
\label{eq:SCET1factorization}
 F(q^2) = H(m_b, E) \, \xi(q^2) \,, 
\end{equation}
with
\begin{equation}
\label{eq:xi_def}
    \xi(q^2) = \frac{1}{2E} \bra{P(p')} J_A(0) \ket{B(v)} \,, 
\end{equation}
and a hard matching coefficient $H(m_b, E)$ encapsulating short-distance fluctuations at the hard scale $E = (m_B^2+m_P^2-q^2)/2m_B \sim m_b$. 

\begin{figure}[t]
	\centering
	\includegraphics[width=0.32\textwidth]{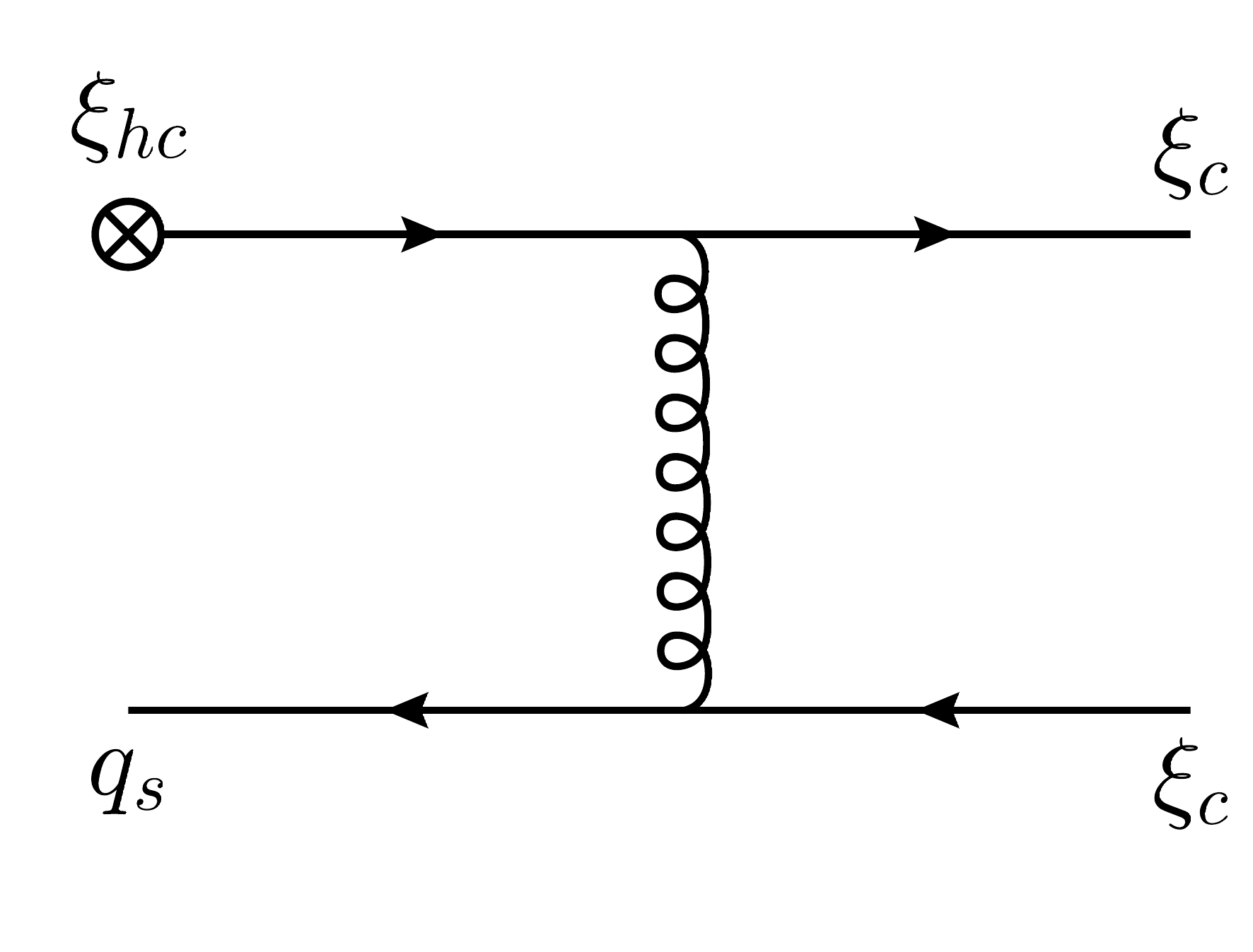}
    \includegraphics[width=0.32\textwidth]{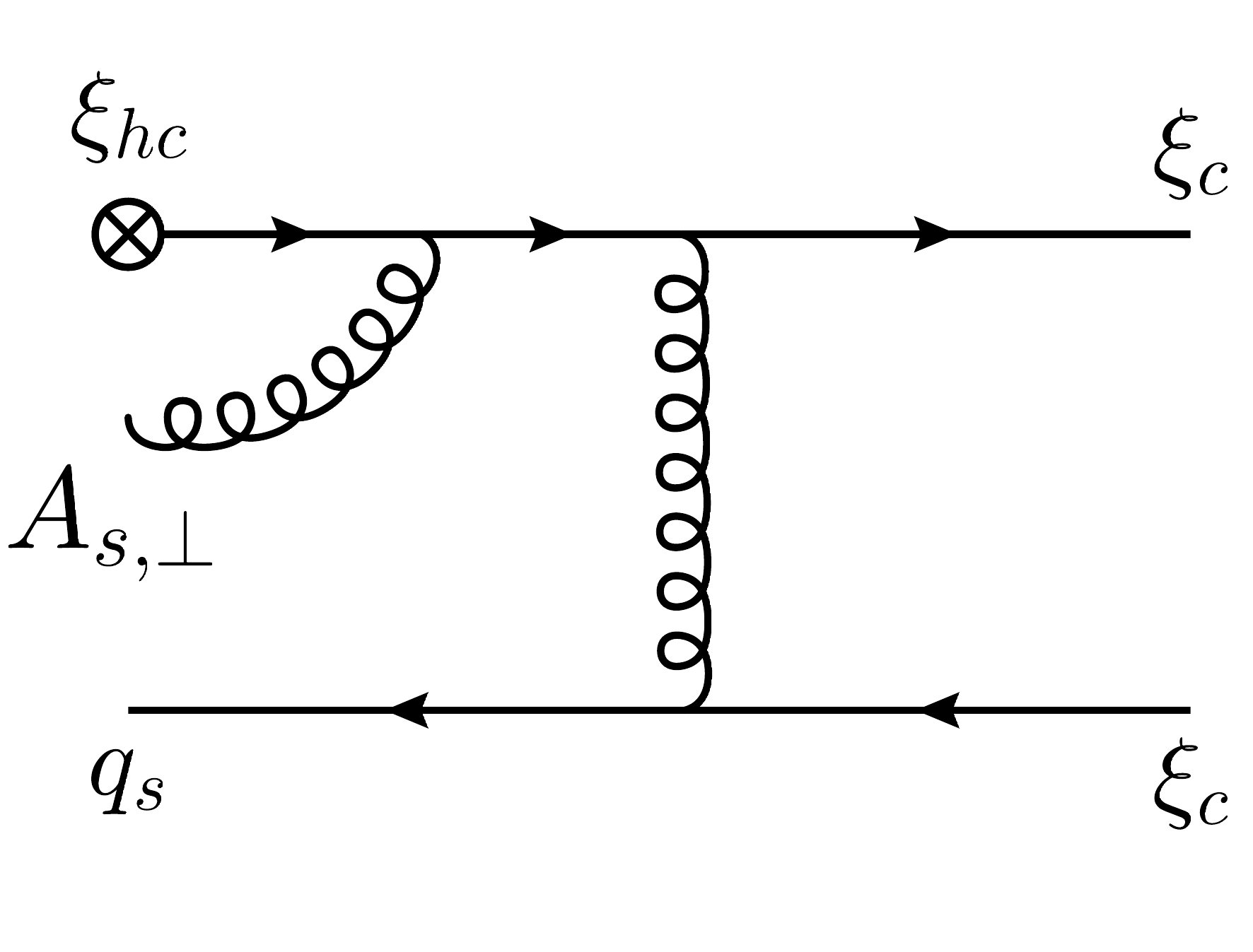}
    \includegraphics[width=0.32\textwidth]{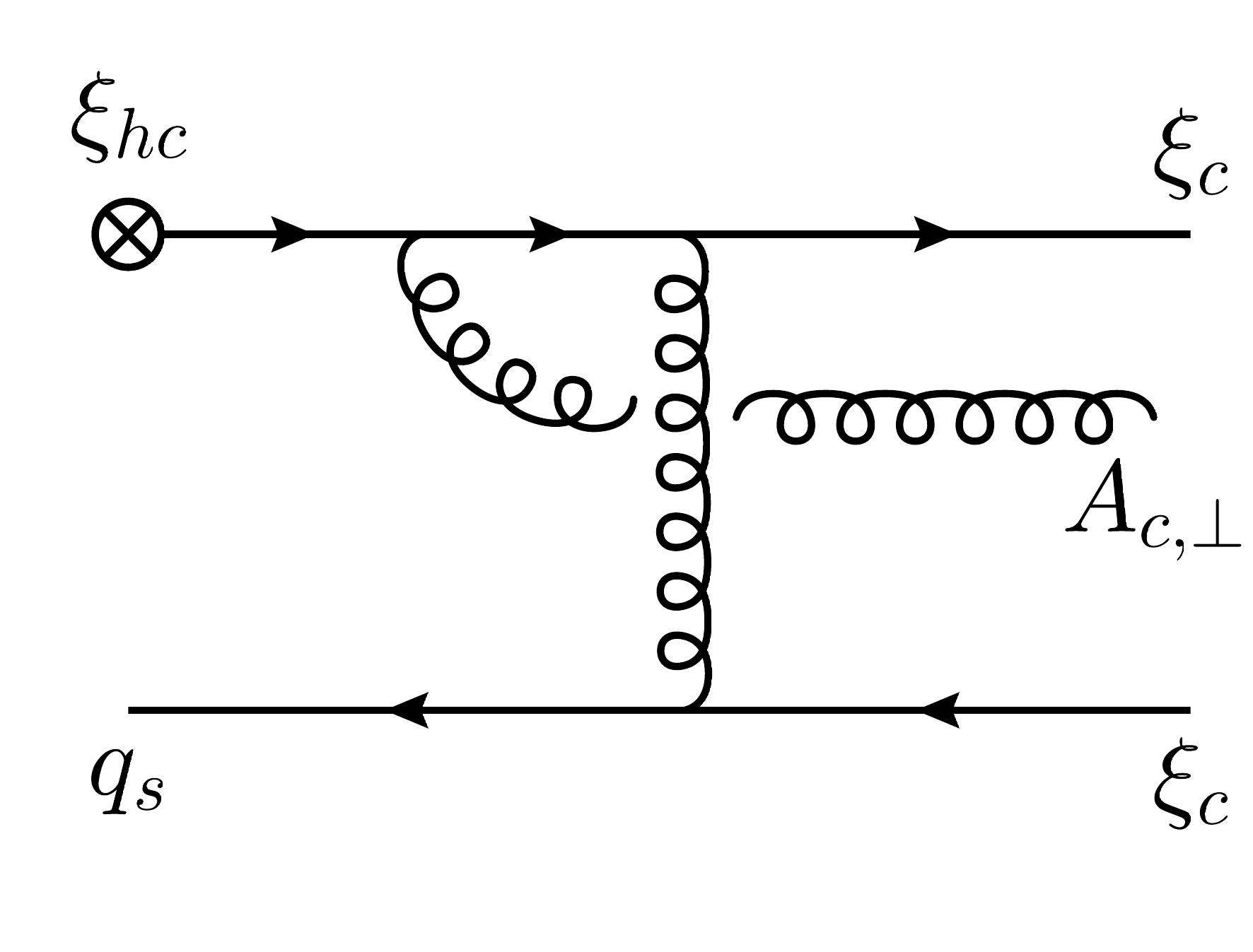}
	\caption{Sample tree-level matching diagrams for the splitting of the hard-collinear \mbox{SCET-1} field $\chi_{hc}$ into soft and collinear SCET-2 fields in light-cone gauge.}
	\label{fig:matching}
\end{figure}

In a second matching step, the time-ordered product $i \int \! d^4 x \, T\{ J_A(0), {\cal L}_{\text{SCET-1}}(x) \}$ with subleading terms in the SCET-1 Lagrangian is matched onto the final effective theory SCET-2. At tree-level, this matching describes a splitting of the hard-collinear quark field $\chi_{hc}$ into soft and collinear quark and gluon fields,
see Fig.~\ref{fig:matching}. In order to have a non-vanishing overlap with the external mesons, the resulting operators must contain at least one soft light-quark field representing the $B$-meson spectator quark, as well as a collinear quark anti-quark pair representing the energetic final-state meson. Such splittings have been studied in~\cite{Beneke:2003pa} at different orders in the power-counting parameter, and it turns out that the first relevant contribution arises at $\mathcal{O}(\lambda^5)$, with $\lambda = (\Lambda_{\rm QCD}/m_b)^{1/2}$.
Quoting only the pieces that are relevant for the soft-overlap form factor, SCET-1 $\to$ SCET-2 matching at tree-level is achieved in light-cone gauge by replacing 
\begin{align}
\label{eq:xi5}
    \xi_{hc}^{(5)} \to - &\frac{1}{i n \cdot \partial} \bigg( (i \slashed{D}_{c, \perp} + g_s \slashed{A}_{s,\perp} + m) \frac{1}{i \bar{n} \cdot \partial} g_s \slashed{A}_{hc,\perp}^{(3)} + g_s \slashed{A}_{hc,\perp}^{(3)} \frac{1}{i \bar{n} \cdot \partial} (i \slashed{D}_{c, \perp} + g_s \slashed{A}_{s,\perp} - m) \bigg) \xi_c \nonumber \\
    - &\frac{1}{i n \cdot \partial} g_s n \cdot A_{hc}^{(5)} \xi_c \,, 
\end{align}
in $J_A$.
Here, $A_{hc,\perp}^{(3) \mu}$ and $n \cdot A_{hc}^{(5)}$ are hard-collinear gluon fields that split into soft and collinear fields according to 
\begin{align}
    A_{hc,\perp}^{\mu,(3)} = g_s t^a \frac{1}{(i \bar{n} \cdot \partial)(i n \cdot \partial)} \big\{\bar{q}_s \gamma_\perp^\mu t^a \xi_c + \text{h.c.} \big\} \,, 
\end{align}
and 
\begin{align}
    n \cdot A_{hc}^{(5)} = &-\frac{2}{(i\bar{n} \cdot \partial)^2} \Bigg\{ i{\cal D}^{\mu_\perp} \big[i \bar{n}\cdot \partial A^{(3)}_{hc, \mu_\perp} \big] - g_s \big[i \bar{n}\cdot \partial A_c^{\mu_\perp}, A^{(3)}_{hc, \mu_\perp} \big] \nonumber \\
    &+2g_s t^a \Big[\bar{\xi}_c t^a \bigg( \frac{\slashed{\bar{n}}}{2} - \frac{1}{in\cdot \partial} g_s \slashed{A}_{c,\perp}\bigg) q_s + \text{h.c.}\Big] \Bigg\} \,, 
\end{align}
where $i {\cal D}^\mu \mathcal{O} = i \partial^\mu \mathcal{O} + g_s [A_c^\mu + A_s^\mu, \mathcal{O}] $. Notice that a non-vanishing quark mass $m\sim \lambda^2 m_b$ has been kept in these expressions, which is precisely what is needed to apply these results to the non-relativistic $B_c \to \eta_c$ form factors in the following section.

These expressions can be cast into a more convenient form by using Fierz identities to group the soft and collinear quark fields into bilinears, projecting out pseudoscalar currents and ``unfixing'' the gauge by re-introducing appropriate soft and collinear Wilson lines. In the limit of massless quarks, this leads to an operator basis for $B \to P$ transitions with four elements~\cite{Lange:2003pk}
\begin{align}
\label{eq:O1-O4}
  \mathcal{O}_1 &= g_s^2 \, \Big[\bar{\chi}_c(0) \frac{\slashed{\bar{n}}}{2} \gamma_5 \chi_c(s \bar{n})\Big] \, \Big[\bar{\mathcal{Q}_s}(\tau n) \frac{\slashed{\bar{n}} \slashed{n}}{4} \gamma_5 \mathcal{H}_v(0)\Big] \,, \nonumber \\
  \mathcal{O}_2 &= g_s^2 \, \Big[\bar{\chi}_c(0) \frac{\slashed{\bar{n}}}{2} \gamma_5 i \slashed{\partial}_\perp \chi_c(s \bar{n})\Big] 			\, \Big[\bar{\mathcal{Q}_s}(\tau n) \frac{\slashed{n}}{2} \gamma_5 \mathcal{H}_v(0)\Big] \,, \nonumber\\
  \mathcal{O}_3 &= g_s^2 \, \Big[\bar{\chi}_c(0) \frac{\slashed{\bar{n}}}{2} \gamma_5 \slashed{\mathcal{A}}_{c,\perp}(r \bar{n}) \chi_c(s \bar{n})\Big] 	\, \Big[\bar{\mathcal{Q}_s}(\tau n) \frac{\slashed{n}}{2} \gamma_5 \mathcal{H}_v(0)\Big] \,, \nonumber\\
  \mathcal{O}_4 &= g_s^2 \, \Big[\bar{\chi}_c(0) \frac{\slashed{\bar{n}}}{2} \gamma_5 \chi_c(s \bar{n})\Big] \, \Big[\bar{\mathcal{Q}_s}(\tau n) \slashed{\mathcal{A}}_{s,\perp}(\sigma n) \frac{\slashed{n}}{2} \gamma_5 \mathcal{H}_v(0)\Big] \,. 
\end{align}
In SCET-2, the gauge-invariant building blocks in the collinear sector read
\begin{align}
 \chi_c = W_c^\dagger \, \xi_c \,, \qquad  \mathcal{A}_{c,\perp} = W_c^\dagger (i D_{c,\perp} W_c) \,, 
\end{align}
and in the soft sector,
\begin{align}
 \mathcal{Q}_s = S_{n}^\dagger \, q_s \,, \qquad \mathcal{H}_v = S_{n}^\dagger \, h_v \,, \qquad \mathcal{A}_{s,\perp} = S_{n}^\dagger (i D_{s,\perp} S_{n}) \,, 
\end{align}
with soft Wilson lines $S_n$ along the $n^\mu$ direction of the energetic final-state $P$ meson. Due to the multipole expansion in SCET \cite{Beneke:2002ni}, soft fields are delocalized along the $n^\mu$ light-cone, and collinear fields along the $\bar{n}^\mu$ light-cone. Note that all Wilson lines in the quark currents combine to straight finite segments connecting the quark and gluon fields. Since in SCET-2, soft-collinear interactions are power-suppressed, the hadronic $B \to P$ matrix elements of the effective operators $\mathcal{O}_i$ factorize into $B \to$ vacuum matrix elements of only soft fields, and vacuum $\to P$ matrix elements of the collinear fields. These matrix elements define LCDA of leading and subleading twist for the heavy and the light meson; more details will be provided in the next section. 

At tree-level, the matching coefficients of the operators $\mathcal{O}_{1-4}$ can either be computed diagrammatically as was done in~\cite{Lange:2003pk}, or they can be directly extracted  from~\eqref{eq:xi5}. The LO coefficient functions in momentum space read
\begin{equation}
\label{eq:D1-D2}
  D_1^{(0)}(u,\omega) = - \frac{C_F}{N_C} \frac{1+\bar{u}}{4 E^2 \bar{u}^2 \omega} \,, \qquad\qquad
  D_2^{(0)}(u,\omega) = - \frac{C_F}{N_C} \frac{1}{4 E^2 u \bar{u}^2 \omega^2 }\,, 
\end{equation}
for the four-quark operators. Here $\omega$ is the light-cone momentum in the $\bar{n}^\mu$ direction of the soft spectator quark in the $B$ meson, and $u$ as well as $\bar{u} = 1-u$ are the light-cone momentum fractions in the $n^\mu$ direction of the quark and anti-quark that form the collinear final-state meson $P$. For the operators involving an additional collinear or soft gluon field, one finds
\begin{align}
\label{eq:D3-D4}
  D_3^{(0)}(\alpha_1,\alpha_2,\alpha_3,\omega) &= \frac{1}{8 N_C E^2 \omega^2 
  \, (\alpha_2+\alpha_3)^2} \left[ -4 C_F + C_A \, \frac{\alpha_2+\alpha_3}{\alpha_2}
  - 2 C_F \, \frac{(\alpha_2+\alpha_3)^2}{\alpha_2(\alpha_1+\alpha_3)} \right]\,, \nonumber\\[0.2cm]
  D_4^{(0)}(u,\omega,\xi) &= \frac{-1}{8 N_C E^2 (\omega+\xi)^2 \bar u^2} \left[ 
  (2 C_F \,\bar{u} - C_A ) \, \frac{\xi}{\omega} + 2 C_F - C_A \right]\,, 
\end{align}
where $\xi$ denotes the soft light-cone momentum of the gluon, and the large collinear momentum of the $P$ meson is distributed for $\mathcal{O}_3$ among the quark ($\alpha_1$), the anti-quark ($\alpha_2$), and the gluon ($\alpha_3$), with $\alpha_1 + \alpha_2 + \alpha_3 = 1$. 

For non-vanishing light-quark masses $m_1$ and $m_2$, the operator basis must be complemented by one additional operator
\begin{equation}
\label{eq:Om}
  \mathcal{O}_m = g_s^2 \, \Big[\bar{\chi}_c(0) \frac{\slashed{\bar{n}}}{2} \gamma_5 \chi_c(s \bar{n})\Big] \, \Big[\bar{\mathcal{Q}_s}(tn) \frac{\slashed{n}}{2} \gamma_5 \mathcal{H}_v(0)\Big] \,. 
\end{equation}
Whereas the operators $\mathcal{O}_{1-4}$ each have a twist-suppression in either the soft or the collinear sector, $\mathcal{O}_m$ is of leading twist in both sectors. However, its matching coefficient 
\begin{equation}
\label{eq:Dm}
  D_m^{(0)}(u,\omega) = \frac{C_F}{N_C} \frac{1}{4 E^2 \bar{u} \, \omega^2}\, \left( m_1 \frac{\bar{u}}{u} +m_2 \frac{u}{\bar{u}} \right) \, ,  
\end{equation} 
is proportional to the quark masses, which gives the correct overall suppression. This matching coefficient has been computed diagrammatically in~\cite{Boer:2018mgl}, but it can also be extracted directly from $\xi_{hc}^{(5)}$. Here the contribution proportional to the mass $m_1$ of the active quark produced in the weak vertex is determined by the two mass terms in the first line of~\eqref{eq:xi5} with $m \to m_1$. The contribution proportional to the spectator-quark mass $m_2$ then arises after using the equations of motion when transverse derivatives act on the soft-quark field $q_s$.

\subsection{Hadronic matrix elements}
\label{subsec:HMEs}

In order to obtain a (bare) factorization formula for the soft-overlap form factor, the meson-to-vacuum matrix elements of the effective quark currents in (\ref{eq:O1-O4}) and \eqref{eq:Om} need to be translated into LCDA. In the soft sector, the leading two-particle 
$B$-meson LCDA $\phi_B^+(\omega)$ is defined as
\begin{equation}
\label{eq:phiplus}
 \bra{0} \bar{\mathcal{Q}}_s(\tau n) \slashed{n} \gamma_5 \mathcal{H}_v(0) \ket{B(v)} = i f_B m_B \int_0^\infty d \omega \, e^{-i\omega \tau} \, \phi_B^+(\omega) \,, 
\end{equation}
where $f_B$ is the $B$-meson decay constant in the static limit. We stress that we work with bare rather than renormalized quantities throughout this section, and the objects in this equation are therefore scale independent. 
The other (subleading) two-particle LCDA $\phi_B^-(\omega)$ is defined via the opposite light-cone projection of the light-quark field, 
\begin{equation}
\label{eq:phiminus}
 \bra{0} \bar{\mathcal{Q}}_s(\tau n) \frac{\slashed{\bar{n}}\slashed{n}}{2} \gamma_5 \mathcal{H}_v(0) \ket{B(v)} = -i f_B m_B \int_0^\infty d \omega \, e^{-i\omega \tau} \, \phi_B^-(\omega) \,. 
\end{equation}
Lastly, the matrix element of the soft current in $\mathcal{O}_4$,
\begin{align}
\label{eq:psiAV}
 \bra{0} \bar{{\cal Q}}_s(\tau n) \slashed{\cal A}_{s,\perp}(\sigma n) \frac{\slashed{n}}{2}\gamma_5 {\cal H}_v(0) \ket{B(v)}
 = i f_B m_B \int_0^\infty d \omega \int_0^\infty \frac{d \xi}{\xi} \, e^{-i\omega\tau} e^{-i \xi \sigma} \phi_{3B}(\omega,\xi) \,, 
\end{align}
defines the relevant three-particle LCDA,
$\phi_{3B} = \Psi_A-\Psi_V$ \cite{Braun:2017liq}. 
These three LCDA are related via an equation of motion, which in $d = 4$ dimensions 
and for a finite spectator quark mass $m_2$ reads~\cite{Bell:2008er}
\begin{align}
\label{eq:softeom1}
  \omega \, \phi_B^-(\omega) - m_2 \, \phi_B^+(\omega) + \int_0^\omega d \eta \left[\phi_B^+(\eta) - \phi_B^-(\eta)\right]
  = 2 \int_0^\omega d \eta \int_{\omega-\eta}^\infty \frac{d\xi}{\xi} \, \frac{\partial}{\partial \xi} \, \phi_{3B}(\eta,\xi) \,. 
\end{align}
Furthermore, from the analysis of their respective RG equations for large scales $\mu \to \infty$, one would expect that the \emph{renormalized} LCDA in the massless case have the following asymptotic
behaviour for small momenta~\cite{Braun:2015pha}
\begin{align}
    \phi_B^+(\omega;\mu) \sim \omega \,, \qquad \phi_B^-(\omega;\mu) \sim 
     \mbox{const.}\,, \qquad  \phi_{3B}(\omega,\xi;\mu) \sim \omega \xi^2 \,. 
\end{align}
However, evaluating the $\omega$-derivative of~\eqref{eq:softeom1} at $\omega = 0$
and using that $\phi_B^+(\omega)$ vanishes for $\omega \to 0$, yields
\begin{align}
  m_2 \, (\phi_B^+)'(0)
  = 2 \int_{0}^\infty \frac{d\xi}{\xi} \, \frac{\partial}{\partial \xi} \, \phi_{3B}(0,\xi) \,, 
\end{align}
which shows that the non-vanishing spectator-quark mass $m_2$ renders the 
LCDA  $\phi_{3B}(\omega,\xi)$ finite as $\omega \to 0$. This has important consequences for the convergence of the inverse moments in the factorization formula discussed below.

In the collinear sector, the leading twist-2 LCDA of a light and energetic  pseudoscalar meson is defined as (see e.g.\ \cite{Ball:1998je})
\begin{align}
  \bra{P(p')} \bar{\chi}_c(0) \slashed{\bar{n}} \gamma_5 \chi_c(s\bar{n}) \ket{0} &= - i \, 2 E \, f_P \int_0^1 d u \, e^{2 i E s \bar{u}} \phi(u) \,. 
\end{align}
The matrix element of the collinear current in $\mathcal{O}_3$ defines a twist-3 three-particle LCDA. 
Expressing the 
non-local QCD operator in the standard definition
\cite{Ball:1998je} in terms of the SCET building blocks,
one obtains
\begin{align}
 \bra{P(p')} \bar{\chi}_c(0) \frac{\slashed{\bar{n}}}{2} \gamma_5 \slashed{\cal A}_{c,\perp}(r \bar{n}) \chi_c(s \bar{n}) \ket{0} 
& = 2i E f_{3P} \int \frac{\mathcal{D}\alpha}{\alpha_3} \, e^{2 i E (\alpha_2 \, s + \alpha_3 \, r )} \, \phi_3(\alpha_1,\alpha_2,\alpha_3) \,,
\end{align}
where we defined the shorthand notation for 
the integration over momentum fractions
\begin{equation}
 \int \mathcal{D} \alpha \equiv \int_0^1 d \alpha_1 \int_0^1 d \alpha_2 \int_0^1 d \alpha_3 \, \delta(1-\alpha_1-\alpha_2-\alpha_3) \,. 
\end{equation}
The collinear current appearing in $\mathcal{O}_2$ can be expressed in terms of $\phi_3$, and the twist-3 two-particle LCDA $\phi_\sigma$ associated with the pseudotensor current. In $d=4$ dimensions,
\begin{align}
  &\bra{P(p')} \bar{\chi}_c(0) \frac{\slashed{\bar{n}}}{2} \gamma_5 i \slashed{\partial}_\perp \chi_c(s \bar{n}) \ket{0} \nonumber \\
	     =\, &2i E \! \int_0^1 d u \, e^{2 i E s \bar{u}} \Bigg\{\frac{f_P \tilde{\mu}_P}{6} \phi_\sigma(u)
	     \! - \! f_{3P} \int_0^u \! d\alpha_1 \int_{u-\alpha_1}^{1-\alpha_1} \! \frac{d\alpha_3}{\alpha_3^2} \, \phi_3(\alpha_1,1-\alpha_1-\alpha_3,\alpha_3) \Bigg\} \,, 
\end{align}
where $\mu_P = m_P^2/(m_1 + m_2)$ is the usual chirally-enhanced mass ratio, and for massive quarks also the combination $\tilde{\mu}_P = \mu_P - (m_1 + m_2)$ appears. Again, the different functions are related via equations of motion, which can be written as~\cite{Ball:1998je}
\begin{align}
\label{eq:collinear_eom1}
  \frac{f_P \tilde{\mu}_P}{6}\phi'_\sigma(u) + (2u-1) \, & f_P \mu_P \, \phi_p(u) - (m_1-m_2) f_P \, \phi(u) \nonumber \\
  =& -2 f_{3P} \int \frac{\mathcal{D}\alpha}{\alpha_3} \, \phi_3(\{ \alpha_i \}) \big( \delta(\alpha_1-u) - \delta(\alpha_2-\bar{u})\big) \,, 
\end{align}
and
\begin{align}
 &\frac{f_P \tilde{\mu}_P}{6} \Big( (2u-1) \phi'_\sigma(u) -4\phi_\sigma(u) \Big) + f_P \mu_P \, \phi_p(u) - (m_1+m_2) f_P \, \phi(u) \nonumber \\
 = \,& 2 f_{3P} \int \frac{\mathcal{D}\alpha}{\alpha_3} \, \phi_3(\{ \alpha_i \}) \left( \delta(\alpha_1-u) + \delta(\alpha_2-\bar{u}) + \frac{2}{\alpha_3} \Big[ \theta(\alpha_1-u) - \theta(\bar{u}-\alpha_2)\Big] \right) \,, 
\end{align}
and also involve the two-particle twist-3 LCDA $\phi_p$ associated with the pseudoscalar quark current. From their respective RG equations one would expect at large scales the asymptotic behavior (see e.g.\ \cite{Ball:1998je}) 
\begin{align}
    \phi(u;\mu) \sim \phi_\sigma(u;\mu) \sim 6u\bar{u} \,, \qquad \phi_p(u;\mu) \sim 1 \,, \qquad \phi_3(\{\alpha_i\};\mu) \sim 
    360 \, \alpha_1 \alpha_2 \alpha_3^2 \,.
\end{align}
Similar to the observation for the three-particle LCDA $\phi_{3B}(\omega,\xi)$ of the $B_c$ meson, the non-vanishing quark masses also alter the functional form of $\phi_3$ for small arguments. Relevant for our discussion is that $\phi_3$ remains finite for $\alpha_2 \to 0$, which can most easily be seen by evaluating~\eqref{eq:collinear_eom1} for $u \to 1$, using that $\tilde{\mu}_P = \mu_P - (m_1 + m_2)$ for the bare parameters.

\subsection{Bare factorization formula}
\label{subsec:LOfactformula}

We are now in the position to express the SCET-1 form factor $\xi(q^2)$ defined in~\eqref{eq:xi_def} in terms of convolution integrals of the various LCDA with the LO matching coefficients $D_{1-4}^{(0)}$ and $D_m^{(0)}$ given in \eqref{eq:D1-D2}, \eqref{eq:D3-D4} and \eqref{eq:Dm} above. The equations of motion allow us to write this result in various ways, and we find the following representation convenient 
\begin{align}
\label{eq:heavytolightFF_naivefacformula_treeHC}
  \xi^{\rm LO}(q^2) =& \, \frac{\alpha_s C_F}{4\pi} \frac{\pi^2 f_{B} f_{P} m_B}{N_c E^2} \bigg\{ \nonumber \\
  &\int_0^\infty \! d \omega \, \int_0^1 \! d u \, 
  \Bigg[ \frac{\phi_B^-(\omega)+\phi_B^+(\omega)}{\omega} \, \frac{1+\bar{u}}{\bar{u}^2} \, \phi(u) -2 \frac{\phi_B^+(\omega)}{\omega} \, \frac{\phi(u)}{\bar{u}} \nonumber \\
  &+ \, \frac{\phi_B^+(\omega)}{\omega^2} \, \Bigg( - m_1 \frac{\phi(u)}{\bar{u}} - 2 m_{2} \frac{\phi(u)}{\bar{u}^2} + 3 \mu_P \frac{\phi_p(u)}{\bar{u}} + \frac{\tilde{\mu}_P}{6} \frac{\phi'_{\sigma}(u)}{\bar{u}} \Bigg) \Bigg] \nonumber \\
  &- \, \frac{C_A-2C_F}{C_F} \, \frac{f_{3P}}{f_P} \, \int_0^\infty \! d \omega \, \frac{\phi_B^+(\omega)}{\omega^2} \, \int {\cal D}\alpha  \, \frac{\phi_3(\{\alpha_i\})}{\alpha_2 \alpha_3 (\alpha_2 + \alpha_3)} \nonumber \\[0.2cm]
  &+ \, \frac{C_A-2 C_F}{C_F} \, \int_0^\infty \! d \omega \, \int_0^\infty \! d \xi \, \frac{\phi_{3B}(\omega,\xi)}{\omega \xi (\omega+\xi)} \, \int_0^1 \! d u \, \frac{\phi(u)}{\bar{u}^2} \bigg\} \,. 
\end{align}
Since $C_A-2C_F = 1/N_c$ the contributions from the three-particle LCDA turn out to be suppressed for large $N_c$ in this form of the result. We emphasize, however, that a word of caution is appropriate when it comes to using the equations of motion, because almost all integrals in the above expression are endpoint-divergent and thus ill-defined, according to the discussion of the asymptotic behavior of the LCDA from the previous section. For the wave functions of the $B_c$ and $\eta_c$ mesons in the non-relativistic approximation, the integrals do not even exist when working in $d \neq 4$ dimensions.
The reason is that the soft and collinear regions of the amplitude develop rapidity divergences, which are by now a well-understood feature of SCET-2 \cite{Becher:2010tm, Becher:2011pf,Chiu:2012ir}. For now, the leading-order factorization formula~\eqref{eq:heavytolightFF_naivefacformula_treeHC} has to be understood in terms of bare quantities, supplemented with suitable rapidity regulators that render all integrals well-defined. We stress that it is in general not guaranteed that the additional regulators will preserve the equations of motion used to derive~\eqref{eq:heavytolightFF_naivefacformula_treeHC}. However, explicit calculation shows that the analytic regulators we choose below do respect the equations of motion in the double-logarithmic approximation, i.e.~for the leading singularities of the bare quantities considered in this work.  
 
\section{Double-logarithms in \texorpdfstring{$B_c \to \eta_c$}{} form factors}
\label{sec:polecancellation}

We now apply the bare factorization formula~\eqref{eq:heavytolightFF_naivefacformula_treeHC} for the SCET-1 form factor $\xi(q^2)$, together with the hard-matching correction encoded in~\eqref{eq:SCET1factorization}, to the soft-overlap form factor for $B_c \to \eta_c$ transitions, in which case the LCDA for both (non-relativistic) mesons can be calculated in perturbation theory. At leading order, the two-particle $B_c$-meson LCDA simply project the light-cone momentum fraction $\omega$ to its on-shell value,
\begin{align}
    \phi_B^{+,(0)}(\omega) = \phi_B^{-,(0)}(\omega) = \delta(\omega - m_2) \,. 
\end{align}
Similarly, the LCDA $\phi(u)$ and $\phi_p(u)$ map the collinear momentum fraction $u$ to the mass ratio $u_0$,
\begin{align}
    \phi^{(0)}(u) = \phi_p^{(0)}(u) = \delta(u-u_0) \,.
\end{align}
The three-particle LCDA $\phi_{3B}(\omega,\xi)$ and $\phi_3(\{\alpha_i\})$ as well as the parameter $\tilde{\mu}_P$ are suppressed by one power of $\alpha_s$.  Note that this implies the vanishing of the hadronic matrix elements of the operators $\mathcal{O}_{2-4}$ at leading order. 
The tree-level result for the form factor $F(\gamma)$ can thus be obtained entirely from $\mathcal{O}_1$ and $\mathcal{O}_m$. The two-particle twist-3 LCDA $\phi_p$ arises in~\eqref{eq:heavytolightFF_naivefacformula_treeHC} because we have used the equations of motion in its derivation. Using that $\mu_P = m_1 + m_2 + \mathcal{O}(\alpha_s)$, and $H(m_b,E_\eta) = 1 + \mathcal{O}(\alpha_s)$, then gives 
\begin{align}
  F^{(0)}(\gamma) = \xi_0 \,
  \frac{2+\bar{u}_0}{\bar{u}_0^3} \,, 
\end{align}
in agreement with~\eqref{eq:xiLO}.

At higher perturbative orders, most of the inverse moments in~\eqref{eq:heavytolightFF_naivefacformula_treeHC} are endpoint-divergent, and hence all LCDA should be considered as bare quantities. Furthermore, it is well-known that in SCET-2 one needs additional rapidity regulators to separate soft and collinear contributions. Despite being a bare factorization formula, it is still useful to independently study and verify the higher-order logarithmic contributions~\eqref{eq:FO} that were predicted in~\cite{Bell:2024bxg} using QCD-based diagrammatic techniques. Instead of the usual RG evolution, large logarithms then arise from the cancellation of singularities between bare quantities, which are accompanied by logarithms in their characteristic virtuality/rapidity. In the following, we identify two different sources of leading singularities: Soft-gluon exchanges result in the usual exponentiated Sudakov factors and are regulated dimensionally.  Soft-quark contributions  manifest in endpoint-divergent convolution integrals of the $B_c$-meson and $\eta_c$-meson LCDA that may require additional rapidity regulators.

\subsection{Next-to-leading order}
\label{subsec:NLOdivergences}

As already mentioned before, radiative corrections receive contributions from hard, hard-collinear, soft and collinear momentum regions, given that no additional regions are introduced through the choice of the analytic regulator.
It is instructive to first analyze how the leading singularities cancel between these regions at the one-loop level, giving rise to the double-logarithmic contribution in the first line of~\eqref{eq:FO}. 

\paragraph{Matching corrections:} 
The matching coefficient $H(m_b, E_\eta)$ in~\eqref{eq:SCET1factorization} arises from cor\-rections to the weak-interaction vertex with virtualities of $\mathcal{O}(m_b^2)$. It is well-known that the leading singularities of the bare coefficient exponentiate to all orders,
\begin{align}
\label{eq:hard_exponentiation}
 H(m_b, E_\eta) \simeq \exp \left\{ -\frac{\alpha_s C_F}{4\pi} \frac{1}{\varepsilon^2} \left( \frac{\mu^2}{4E_\eta^2} \right)^\varepsilon \right\} = 1 -\frac{\alpha_s C_F}{4\pi} \frac{1}{\varepsilon^2}\left( \frac{\mu^2}{4E_\eta^2} \right)^\varepsilon + \mathcal{O}(\alpha_s^2) \,.    
\end{align}
Thus, at NLO the leading singularities of $F(\gamma)$ from hard contributions are given by
\begin{align}
\label{eq:NLOhardpoles}
    F^{(1)}_h(\gamma) \simeq -\frac{C_F}{\varepsilon^2} \left( \frac{\mu^2}{4E_\eta^2} \right)^\varepsilon F^{(0)}(\gamma) \,. 
\end{align}
The $\varepsilon$-expansion of $H(m_b, E_\eta)$  generates logarithms in the ratio of the factorization scale $\mu^2$ and the intrinsic hard scale of the process $\mu_h^2 \sim m_b^2 \sim 4E_\eta^2$.

Radiative corrections from the hard-collinear region are more involved. At $\mathcal{O}(\alpha_s)$, only the corrections to the coefficient functions $D_1$ and $D_m$ are relevant for the double logarithms, as the leading-order matrix elements of the operators $\mathcal{O}_{2-4}$ vanish. These require to compute one-loop corrections to the left diagram in Fig.~\ref{fig:matching}.
The leading singularities are found to be (see also~\cite{Boer:2018mgl})
\begin{align}
\label{eq:D1NLO}
 D_1^{(1)}(u,\omega) \simeq - \frac{C_F}{N_C} \frac{1 + \bar{u}}{4 E^2 \bar{u}^2 \omega} \frac{4C_F}{\varepsilon^2} \left( \frac{\mu^2}{2 \omega E_\eta} \right)^\varepsilon \,, 
\end{align}
and
\begin{align}
\label{eq:DmNLO}
 D_m^{(1)}(u,\omega) \simeq \frac{C_F}{N_C} \frac{1}{4 E^2 \bar{u} \, \omega^2}\, \frac{4}{\varepsilon^2} \left( \frac{\mu^2}{2 \omega E_\eta} \right)^\varepsilon \left\{ \left( m_1 \frac{\bar{u}}{u} + m_2 \frac{3}{2\bar{u}} \right) C_F - m_2 \frac{C_A}{4\bar{u}} \right\} \,. 
\end{align}
Evaluating the hadronic matrix elements of $\mathcal{O}_1$ and $\mathcal{O}_m$ at tree-level then results in
\begin{align}
\label{eq:NLOhcpoles}
 F^{(1)}_{hc}(\gamma) \simeq \frac{2C_F}{\varepsilon^2} \left( \frac{\mu^2}{2 m_\eta E_\eta} \right)^\varepsilon F^{(0)}(\gamma) 
+ \frac{\xi_0}{\varepsilon^2} \left( \frac{\mu^2}{2 m_\eta E_\eta} \right)^\varepsilon \bigg(
 6C_F \, \frac{1+\bar{u}_0}{\bar{u}_0^3}
- \frac{C_A}{\bar{u}_0^3} \bigg) \,, 
\end{align}
for the leading singularities of the hard-collinear region. Again, the $\varepsilon$-expansion generates logarithms which now involve the typical hard-collinear scale $\mu_{hc}^2 \sim 2m_\eta E_\eta$, where we have neglected quantities of $\mathcal{O}(1)$ in the scale ratios. For this contribution, the leading singularities are clearly more intricate than a simple exponential multiplying the leading-order expression, as both, the color structures and the dependence on the quark-mass ratios differ from the LO result in~\eqref{eq:xiLO}.

\paragraph{Hadronic matrix elements:} 
The double poles in the dimensional regulator $\varepsilon$ from the hard and hard-collinear matching contributions need to cancel against singularities that arise from the soft and collinear loops. The latter can be obtained from the bare factorization formula~\eqref{eq:heavytolightFF_naivefacformula_treeHC} by evaluating the LCDA to next-to-leading order in perturbation theory. Here, divergences arise from two sources: The LCDA themselves feature the usual UV singularities, but as we will see also their endpoint-divergent inverse moments provide an additional source of double-logarithmic corrections.

Starting with the purely collinear contribution, we first note that all $\eta_c$-meson LCDA are intrinsically single-logarithmic objects that obey Brodsky-Lepage-like evolution equations~\cite{Lepage:1979zb,Lepage:1980fj,Efremov:1979qk}. Double poles thus only arise from the endpoint-divergent convolutions in~\eqref{eq:heavytolightFF_naivefacformula_treeHC} as $u \to 1$. In this approximation, the inverse moment of the three-particle LCDA $\phi_3$ can be replaced by means of the equations of motion via
\begin{align}
 2 f_{3P}\, \int {\cal D}\alpha \, \frac{\phi_3(\{\alpha_i\})}{\alpha_2 \alpha_3 (\alpha_2 + \alpha_3)} \simeq \, f_P \int_0^1 \! d u \, 
  \Bigg( m_2 \frac{\phi(u)}{\bar{u}^2} - \mu_P \frac{\phi_p(u)}{\bar{u}} - \frac{\tilde{\mu}_P}{6} \frac{\phi'_{\sigma}(u)}{\bar{u}} \Bigg) \,. 
\end{align}
After inserting the LO $B_c$-meson LCDA in \eqref{eq:heavytolightFF_naivefacformula_treeHC}, the purely collinear contribution can be expressed entirely through two-particle LCDA of the $\eta_c$ meson,
\begin{align}
\label{eq:Fcoll}
 \bar{u}_0^2 \frac{F(\gamma)\big\vert_{\rm coll}}{\xi_0} \simeq & - \frac{u_0}{\bar{u}_0} 
  + \bar{u}_0 \frac{2C_F - C_A}{2C_F} \int_0^1 \! \frac{d u}{\bar{u}^2} \phi(u) 
  + \frac{2\mu_P}{m_\eta} \int_0^1 \! \frac{d u}{\bar{u}} \phi_p(u)
  + \frac{C_A}{2C_F} \int_0^1 \! \frac{d u}{\bar{u}} \phi_{p+\sigma'}(u) \,. 
\end{align}
Here, we abbreviated $m_\eta \phi_{p+\sigma'} \equiv \mu_P \phi_p + \frac{\tilde{\mu}_P}{6} \phi'_\sigma$, and the first term on the right-hand side of the equation originates from an endpoint-finite moment of $\phi$, which only contributes at tree-level in the double-logarithmic approximation. All relevant collinear LCDA have been computed in~\cite{Bell:2008er} at NLO. Expanding these functions near the endpoint $u \to 1$, and keeping only the UV-divergent piece, yields
\begin{align}
    \phi^{(1)}(u \to 1) &\simeq \frac{2C_F}{\varepsilon} \left(\frac{\mu^2}{m_2^2}\right)^\varepsilon \frac{1+\bar{u}_0}{\bar{u}_0^2} \, \bar{u} \,,\nonumber \\
    \phi_p^{(1)}(u \to 1) &\simeq \frac{2C_F}{\varepsilon} \left(\frac{\mu^2}{m_2^2}\right)^\varepsilon \frac{1+\bar{u}_0}{\bar{u}_0} \,, \nonumber \\
    \phi_{p+\sigma'}^{(1)}(u \to 1) &\simeq \frac{2C_F}{\varepsilon} \left(\frac{\mu^2}{m_2^2}\right)^\varepsilon \,, 
\end{align}
which shows that the integrals in~\eqref{eq:Fcoll} are ill-defined even in $d$ space-time dimensions. In order to regularize these rapidity divergences, we modify each integration measure according to~\cite{Becher:2011dz}
\begin{equation}
\label{eq:regulator}
    d^d k \to d^d k \left( \frac{\nu}{k_-} \right)^\alpha \,, 
\end{equation}
where $k^\mu$ is the momentum of the spectator quark that flows out of the hard-collinear interaction vertex, i.e. $k_- = 2 \bar{u} E_\eta$.
At NLO, all endpoint divergences in the collinear sector then show up as poles in $\alpha$. 
Employing this prescription, the divergent inverse moments evaluate to 
\begin{align}
    \int_0^1 \! \frac{d u}{\bar{u}^2} \phi^{(1),\text{reg}}(u) &\simeq \frac{-2C_F}{\alpha\varepsilon} \left(\frac{\mu^2}{m_\eta^2}\right)^\varepsilon \left( \frac{\nu}{2E_\eta} \right)^\alpha \frac{1+\bar{u}_0}{\bar{u}_0^2}\,, \nonumber\\
    \frac{\mu_P}{m_\eta} \int_0^1 \! \frac{d u}{\bar{u}} \phi_p^{(1),\text{reg}}(u) &\simeq \frac{-2C_F}{\alpha\varepsilon} \left(\frac{\mu^2}{m_\eta^2}\right)^\varepsilon \left( \frac{\nu}{2E_\eta} \right)^\alpha \frac{1+\bar{u}_0}{\bar{u}_0}\,, \nonumber\\
    \int_0^1 \! \frac{d u}{\bar{u}} \phi^{(1),\text{reg}}_{p+\sigma'}(u) &\simeq \frac{-2C_F}{\alpha\varepsilon} \left(\frac{\mu^2}{m_\eta^2}\right)^\varepsilon \left( \frac{\nu}{2E_\eta} \right)^\alpha \,. 
    \label{eq:NLO:collinearmoments}
\end{align}
According to~\eqref{eq:Fcoll} the double poles in the collinear sector then add up to
\begin{align}
\label{eq:NLOcollinearpoles}
 F^{(1)}_c(\gamma) \simeq \frac{-\xi_0}{\alpha\varepsilon} \left(\frac{\mu^2}{m_\eta^2}\right)^\varepsilon \left( \frac{\nu}{2E_\eta} \right)^\alpha \bigg(6C_F \frac{1+\bar{u}_0}{\bar{u}_0^3} - C_A \, \frac{1}{\bar{u}_0^3} \bigg) \,, 
\end{align}
and we note that the natural rapidity scale in the collinear sector is large, $\nu \sim 2 E_\eta$, whereas the virtuality of this mode is small $\mu^2 \sim m_\eta^2$.

Moving on to the soft sector, we first note that the $B_c$-meson LCDA are double-logarithmic quantities that obey Lange-Neubert-like evolution~\cite{Lange:2003ff}. The reason is that the soft operators in HQET contain a time-like Wilson line (the heavy-quark field) in the $v^\mu$ direction and a light-like Wilson line in the $n^\mu$ direction, which meet at the hard-interaction point, i.e. the weak vertex, and form a cusp.  In contrast to the collinear sector, the (bare) expressions of the $B_c$-meson LCDA thus contain double poles in $\varepsilon$. In addition, endpoint-divergent convolutions generate a second source of double poles in the limit $\omega\to0$, which require a rapidity regulator. Following similar steps as before, we first use that 
\begin{equation}
\frac{1}{\omega\xi(\omega+\xi)} = \frac{1}{\omega(\omega+\xi)^2} + \frac{1}{\xi(\omega+\xi)^2} \,.
\end{equation}
The integral of the last term with the three-particle LCDA $\phi_{3B}$ is endpoint-finite and $\alpha_s$ suppressed, and is thus irrelevant for the double logarithms, see the one-loop expression in~\eqref{eq:phi3BNLO}. The equation of motion~\eqref{eq:softeom1} then allows us to trade the contribution from $\phi_{3B}$ with $\phi_B^+$ via
\begin{align}
\label{eq:softeom}
   2 \int_0^\infty \! d \omega \, \int_0^\infty \! d \xi \, \frac{\phi_{3B}(\omega,\xi)}{\omega (\omega+\xi)^2} \simeq \int_0^\infty \frac{d \omega}{\omega} \phi_B^+(\omega) \left(1 - \frac{m_2}{\omega} \right) \,.  
\end{align}
Replacing further the $\eta_c$-meson LCDA in~\eqref{eq:heavytolightFF_naivefacformula_treeHC} with their LO expressions yields 
\begin{align}
\label{eq:purelysoftregion}
  \frac{F(\gamma)\big\vert_{\text{soft}}}{\xi_0}
  \simeq &\int_0^\infty \!\! d \omega \, \frac{m_2}{\omega} \Bigg\{ \!\frac{1+\bar{u}_0}{\bar{u}_0^3} \bigg(\phi_B^-(\omega)
  + \frac{m_2}{\omega} \phi_B^+(\omega) \bigg) 
  - \frac{\phi_B^+(\omega)}{\bar{u}_0^2} 
  + \frac{C_A}{2C_F \bar{u}_0^3} \, \phi_B^+(\omega) \, \bigg( 1 - \frac{m_2}{\omega} \!\bigg) \!\Bigg\}  
\end{align}
for the purely soft contribution to the form factor. Here, we write each inverse power of $\omega$ with a factor $m_2 = \bar{u}_0 m_\eta$, such that the appearing soft integrals are dimensionless and independent of powers of the quark-mass ratio $\bar{u}_0$.

In contrast to the collinear moments, the NLO soft contribution cannot be evaluated by simply using the one-loop expressions for the $B_c$-meson LCDA derived in~\cite{Bell:2008er}, supplemented with a regulator $\sim \omega^\alpha$.
The reason is that the prescription~\eqref{eq:regulator} must be used consistently at the level of Feynman integrals in each momentum region. 
It turns out, that the soft integrals yield double singularities of the form
\begin{equation}
\label{eq:alphaexpansion}
    \frac{\Gamma(\alpha+\varepsilon)}{\alpha} \simeq \frac{1}{\alpha\varepsilon} - \frac{1}{\varepsilon^2} \,, 
\end{equation}
which need to be expanded first in $\alpha$ and then in $\varepsilon$. In comparison to     \eqref{eq:NLO:collinearmoments}, our regulator choice thus guarantees that at NLO all double poles in $\varepsilon$ associated with endpoint-divergent convolution integrals are part of the soft sector. 
Explicit calculation gives 
\begin{align}
    \int_0^\infty \! d \omega \, \frac{m_2^2}{\omega^2} \, \phi_{B,\text{reg}}^{+,(1)}(\omega) &\simeq 2 C_F \left(\frac{1}{\alpha\varepsilon} - \frac{1}{\varepsilon^2} \right) \left(\frac{\mu^2}{m_2^2} \right)^\varepsilon \left( \frac{\nu}{m_2} \right)^\alpha - \frac{C_F}{\varepsilon^2} \left(\frac{\mu^2}{m_2^2}\right)^\varepsilon\,, \nonumber\\
    \int_0^\infty \! d \omega \, \frac{m_2}{\omega} \, \phi_{B,\text{reg}}^{-,(1)}(\omega) &\simeq 4 C_F \left(\frac{1}{\alpha\varepsilon} - \frac{1}{\varepsilon^2} \right) \left(\frac{\mu^2}{m_2^2} \right)^\varepsilon \left( \frac{\nu}{m_2} \right)^\alpha - \frac{C_F}{\varepsilon^2} \left(\frac{\mu^2}{m_2^2}\right)^\varepsilon\,,  
     \label{eq:NLO:softmoments}
\end{align}
where in both expressions the first term arises from the endpoint singularity, and the second term is the cusp contribution. Note that the soft sector is characterized by a small rapidity, $\nu \sim m_2$, and a small virtuality, $\mu^2 \sim m_2^2$. Due to the cusp in the soft operators, the (endpoint-finite) first inverse moment of the leading-twist LCDA is also double-logarithmic,
\begin{equation}
\label{eq:NLO:lamB}
  \int_0^\infty \! d \omega \, \frac{m_2}{\omega} \, \phi_B^{+,(1)}(\omega) \simeq - \frac{C_F}{\varepsilon^2} \left( \frac{\mu^2}{m_2^2} \right)^\varepsilon \,. 
\end{equation}
Adding up the terms in~\eqref{eq:purelysoftregion} then yields for all double poles in the NLO soft contribution
\begin{align}
\label{eq:NLOsoftpoles}
  F^{(1)}_s(\gamma)
  \simeq \xi_0 \left(\frac{\mu^2}{m_\eta^2} \right)^\varepsilon \left( \frac{\nu}{m_\eta} \right)^\alpha \left( 6 C_F \frac{1+\bar{u}_0}{\bar{u}_0^3} - \frac{C_A}{\bar{u}_0^3}\right) \left(\frac{1}{\alpha\varepsilon} - \frac{1}{\varepsilon^2} \right) - \frac{C_F}{\varepsilon^2} \left(\frac{\mu^2}{m_\eta^2}\right)^\varepsilon F^{(0)}(\gamma) \,, 
\end{align}
where the term proportional to the LO form factor $F^{(0)}(\gamma)$ is the cusp contribution, and we again ignored parametrically small logarithms of the mass ratios $u_0$ and $\bar{u}_0$ by replacing the spectator-quark mass $m_2$ with the meson mass $m_\eta$ in the virtuality and rapidity ratios. 

Summing up the terms from all four regions -- hard~\eqref{eq:NLOhardpoles}, hard-collinear~\eqref{eq:NLOhcpoles}, collinear \eqref{eq:NLOcollinearpoles} and soft~\eqref{eq:NLOsoftpoles} -- we observe that all singularities in $\alpha$ and $\varepsilon$ cancel, and the associated natural rapidity and virtuality scales combine to large double logarithms,
\begin{align}
    F^{(1)}(\gamma) \simeq \xi_0 \left( 6 C_F \frac{1+\bar{u}_0}{\bar{u}_0^3} - \frac{C_A}{\bar{u}_0^3}\right) \frac{L^2}{2} -C_F L^2 \, F^{(0)}(\gamma) = \xi_0 L^2 \left( C_F \frac{1+2\bar{u}_0}{\bar{u}_0^3} - \frac{C_A}{2\bar{u}_0^3} \right) \,,
\end{align}
which is the same result as given in the first line of~\eqref{eq:FO}. In summary, this analysis reflects the two mechanisms identified in~\cite{Bell:2024bxg} relevant for the leading double logarithms of the soft-overlap form factor. First, \emph{soft-gluon configurations} lead to standard exponentiated Sudakov factors. At NLO these are associated with double poles in the dimensional regulator in the hard, hard-collinear and soft region, and they have their origin in a cusp at the heavy-to-light interaction vertex. Second, \emph{soft-quark configurations} are connected to endpoint singularities of the heavy- and light-meson LCDA. The associated singularities show up as poles in the dimensional regulator and the rapidity regulator in the two low-energy regions, i.e. soft and collinear. In addition, also the hard-collinear region carries information about these configurations in the form of double poles in $\varepsilon$. In the sum of all four regions, the singularities cancel and the logarithms of the different rapidities and virtualities combine to the double-logarithmic corrections in $F(\gamma)$.

\subsection{Next-to-next-to-leading order}
\label{subsec:NNLO}

At higher loop orders, we expect that the interplay of soft-gluon and soft-quark corrections leads to the recursive integral equations in~\eqref{eq:finalintegralequations}. In view of the complexity of such a SCET analysis, however, we refrain from computing the leading singularities of all contributing regions beyond one-loop order in this work. We will rather turn the argument around in the following, and \emph{assume} that no new momentum regions show up at higher orders in perturbation theory. Therefore the bare factorization formula \eqref{eq:heavytolightFF_naivefacformula_treeHC} in SCET, with all singularities either regularized dimensionally or by the rapidity regulator, correctly reproduces the dynamics at every fixed order. The various poles in the regulators are then bound to cancel in the sum of all regions, and their interplay generates large logarithmic corrections to the form factor, similar to what we have seen at NLO in the previous section. Once we assume that this pole cancellation will happen, one may ask how many ingredients are needed to predict the double logarithms at the two-loop level and beyond. 

Before entering the technical details of such an approach, we stress that this method is independent from the setup we used previously in~\cite{Bell:2024bxg}. In that work, we extracted the double logarithms diagrammatically, starting from a single (non-standard) overlap region and furnishing the relevant integrations over the longitudinal momentum components with physical cutoffs. In contrast, here we assume that the double logarithms are generated through an interplay between the standard SCET regions, which are defined without any cutoffs, but provided with appropriate (analytic) regulators. In particular, the overlap region considered in~\cite{Bell:2024bxg} is scaleless and vanishes in the EFT approach.

In order to illustrate the idea and to set up the notation, we find it instructive to first review the NLO calculation along these lines. Without any specific information about the underlying  SCET operators and their hadronic matrix elements, we now assume that the double logarithms are generated through an interplay of four regions, 
\begin{align}
    F^{(1)}(\gamma) =  F^{(1)}_h(\gamma) + F^{(1)}_{j}(\gamma) + F^{(1)}_c(\gamma) + F^{(1)}_s(\gamma) \,,
    \label{eq:NLO:sumofallregions}
\end{align}
which in general have the following singularity structure at the double-logarithmic level 
\begin{align}
    F^{(1)}_h(\gamma) &\simeq \frac{h_{20}}{\varepsilon^2}  \left( \frac{\mu^2}{4E_\eta^2} \right)^\varepsilon F^{(0)}(\gamma) \,, \cr
    F^{(1)}_{j}(\gamma) &\simeq \frac{j_{20}}{\varepsilon^2}  \left( \frac{\mu^2}{2m_\eta E_\eta} \right)^\varepsilon \,, \cr 
    F^{(1)}_c(\gamma) &\simeq \bigg\{\frac{c_{20}}{\varepsilon^2} + \frac{c_{11}}{\alpha\varepsilon} \bigg\} \left( \frac{\mu^2}{m_\eta^2} \right)^\varepsilon \left( \frac{\nu}{2 E_\eta} \right)^\alpha \,, \cr 
    F^{(1)}_s(\gamma) &\simeq \bigg\{\frac{s_{20}}{\varepsilon^2} + \frac{s_{11}}{\alpha\varepsilon} \bigg\} \left( \frac{\mu^2}{m_\eta^2} \right)^\varepsilon \left( \frac{\nu}{m_\eta} \right)^\alpha \,, 
    \label{eq:NLO:ansatz}
\end{align}
where we used that the hard contribution factorizes to the tree-level result according to \eqref{eq:SCET1factorization}. Moreover, we denoted the hard-collinear modes more concisely as jet modes for notational purposes, and we used that the rapidity divergences can only appear in the low-scale contributions, because of a double counting of soft and collinear fluctuations with the same virtuality. Most importantly, each of the leading singularities generates a tower of logarithms in the corresponding virtuality and rapidity scales in an expansion in the two regulators. 

At NLO the ansatz \eqref{eq:NLO:ansatz} contains six coefficients that in general depend on the quark mass ratio $\bar{u}_0$, and which could easily be read off from the expressions provided in the previous section. Here, we instead assume that these coefficients are unknown, and we demand that the various poles must cancel in the sum of all regions. This pole-cancellation argument then yields non-trivial relations among the coefficients. Explicitly, we find that only two coefficients are needed to predict the double logarithms at one loop, which we may choose to be $h_{20}$ and $j_{20}$, or we may substitute the latter by $c_{11}$. In terms of these coefficients, we find that the one-loop double logarithms can be written in the form  
\begin{align}
    F^{(1)}(\gamma) \simeq \bigg(2 h_{20} F^{(0)}(\gamma) + \frac{j_{20}}{2}\bigg) L^2
    = \bigg( h_{20} F^{(0)}(\gamma) - \frac{c_{11}}{2}\bigg) L^2\,.
\end{align}
One easily verifies that this relation reproduces the known result in~\eqref{eq:FO} for the coefficients provided in \eqref{eq:NLOhardpoles} and \eqref{eq:NLOhcpoles} or \eqref{eq:NLOcollinearpoles} above that are also collected in App.~\ref{app:coefficients} for convenience. We thus conclude that the one-loop logarithms can be entirely predicted from the matching corrections $h_{20}$ and $j_{20}$ alone. In view of the two underlying dynamical mechanisms, one may alternatively choose $h_{20}$ and $c_{11}$ to quantify the soft-gluon and soft-quark corrections.

Proceeding next to NNLO, the form factor receives contributions from ten regions. First of all, there are four ``pure'' regions, which can be parametrized in analogy to \eqref{eq:NLO:ansatz}, except that the low-scale regions may now contain up to two poles in the rapidity regulator, 
\begin{align}
    F^{(2)}_{hh}(\gamma) &\simeq \frac{h_{40}}{\varepsilon^4}  \left( \frac{\mu^2}{4E_\eta^2} \right)^{2\varepsilon} F^{(0)}(\gamma) \,, \cr
    F^{(2)}_{jj}(\gamma) &\simeq \frac{j_{40}}{\varepsilon^4}  \left( \frac{\mu^2}{2m_\eta E_\eta} \right)^{2\varepsilon} \,, \cr 
    F^{(2)}_{cc}(\gamma) &\simeq \bigg\{\frac{c_{40}}{\varepsilon^4} + \frac{c_{31}}{\alpha\varepsilon^3}+ \frac{c_{22}}{\alpha^2\varepsilon^2} \bigg\} \left( \frac{\mu^2}{m_\eta^2} \right)^{2\varepsilon} \left( \frac{\nu}{2 E_\eta} \right)^{2\alpha} \,, \cr 
    F^{(2)}_{ss}(\gamma) &\simeq \bigg\{\frac{s_{40}}{\varepsilon^4} + \frac{s_{31}}{\alpha\varepsilon^3} + \frac{s_{22}}{\alpha^2\varepsilon^2} \bigg\} \left( \frac{\mu^2}{m_\eta^2} \right)^{2\varepsilon} \left( \frac{\nu}{m_\eta} \right)^{2\alpha} \,, 
    \label{eq:NNLOpure:ansatz}
\end{align}
which introduces eight unknown coefficients. In addition, there are six mixed regions, of which the ones that involve the hard region are simply given by products of the one-loop coefficients from \eqref{eq:NLO:ansatz}, because of the multiplicative structure of the SCET-1 factorization theorem in~\eqref{eq:SCET1factorization}. For the remaining three regions, we make the ansatz 
\begin{align}
    F^{(2)}_{jc}(\gamma) &\simeq \bigg\{\frac{(jc)_{40}}{\varepsilon^4} + \frac{(jc)_{31}}{\alpha\varepsilon^3}+ \frac{(jc)_{22}}{\alpha^2\varepsilon^2} \bigg\} \left( \frac{\mu^2}{2m_\eta E_\eta} \right)^{\varepsilon}  \left( \frac{\mu^2}{m_\eta^2} \right)^{\varepsilon} \left( \frac{\nu}{2 E_\eta} \right)^{\alpha}\,, \cr 
    F^{(2)}_{js}(\gamma) &\simeq \bigg\{\frac{(js)_{40}}{\varepsilon^4} + \frac{(js)_{31}}{\alpha\varepsilon^3}+ \frac{(js)_{22}}{\alpha^2\varepsilon^2} \bigg\} \left( \frac{\mu^2}{2m_\eta E_\eta} \right)^{\varepsilon}  \left( \frac{\mu^2}{m_\eta^2} \right)^{\varepsilon} \left( \frac{\nu}{m_\eta} \right)^{\alpha}\,, \cr  
    F^{(2)}_{cs}(\gamma) &\simeq \bigg\{\frac{(cs)_{40}}{\varepsilon^4} + \frac{(cs)_{31}}{\alpha\varepsilon^3} + \frac{(cs)_{22}}{\alpha^2\varepsilon^2} \bigg\} \left( \frac{\mu^2}{m_\eta^2} \right)^{2\varepsilon} \left( \frac{\nu}{2 E_\eta} \right)^{\alpha}\left( \frac{\nu}{m_\eta} \right)^{\alpha} \,, 
    \label{eq:NNLOmixed:ansatz}
\end{align}
which introduces nine more coefficients.

Working through the pole-cancellation arguments, we then find that there are only three genuine two-loop coefficients needed to predict the double logarithms at this order. For reasons that will become clearer below, we choose the set $\{h_{40}, j_{40}, (cs)_{22}\}$, in terms of which the leading two-loop logarithms can be reconstructed via 
\begin{align}
\label{eq:F2:coefficients}
    F^{(2)}(\gamma) \simeq \bigg(
    6 h_{40} F^{(0)}(\gamma) + \frac{j_{40}}{6} - \frac{cs_{22}}{24} - \frac76 (h_{20})^2 F^{(0)}(\gamma) + \frac56 h_{20}\, j_{20}\bigg) L^4\,.
\end{align}
Two of the novel coefficients can easily be determined. First, the leading logarithms in the hard function -- associated with standard soft-gluon dynamics -- are known to exponentiate, i.e.~one has $h_{40}=(h_{20})^2/2$. Moreover, the coefficient $(cs)_{22}$ involves only ingredients that were given in the previous sections, namely the LO factorization formula~\eqref{eq:heavytolightFF_naivefacformula_treeHC} as well as the endpoint-divergent one-loop moments~\eqref{eq:NLO:collinearmoments} and~\eqref{eq:NLO:softmoments}. From these expressions, we obtain (see also (D.18) in~\cite{Boer:2018mgl}), 
\begin{align}
\label{eq:cs22}
    (cs)_{22}=
 \xi_0 C_F \bigg(-16C_F \frac{1+\bar{u}_0}{\bar{u}_0^3} +2 C_A \, \frac{2+\bar{u}_0}{\bar{u}_0^3} \bigg) \,. 
\end{align}
This then leaves one unknown two-loop coefficient, for which we have chosen $j_{40}$, since its determination is independent of any ambiguities associated with the rapidity regulator, and it can therefore be computed with standard multi-loop techniques. We will report on the details of this two-loop calculation in Sec.~\ref{sec:FOcalculation}.

\subsection{Next-to-next-to-next-to-leading order}
\label{subsec:N3LO}

We will now extend this approach to one order higher, i.e.~to N$^3$LO. Indeed, the intricate structure of the integral equations~\eqref{eq:finalintegralequations} exhibits certain features only at this order, such as the exponentiation of soft-gluon corrections within the soft-quark integrals. In any case, an independent derivation of the three-loop logarithms using SCET methods constitutes a powerful validation of the all-order result.

At N$^3$LO the number of independent regions and the corresponding coefficients of the leading singularities proliferates. As the strategy has already been discussed in detail above, we omit the corresponding discussion here. Also, the notation we use for the coefficients is identical to the previous section, e.g.~$(jjs)_{51}$ refers to the coefficient of the $1/(\alpha\varepsilon^5)$ pole in the mixed (hard-collinear)-(hard-collinear)-soft region. We then start from four pure regions with 10 coefficients in analogy to~\eqref{eq:NNLOpure:ansatz}, nine multiplicative contributions that involve the hard region and that depend only on lower-order coefficients, as well as seven non-trivial mixed regions that give rise to another 28 coefficients similar to~\eqref{eq:NNLOmixed:ansatz}. In total, these are thus 38 unknown three-loop coefficients.

Similar to~\eqref{eq:F2:coefficients}, the pole-cancellation conditions imply that the coefficient of the leading double-logarithmic contribution at N$^3$LO can be expressed in terms of a much smaller set of three-loop coefficients, in addition to lower-order coefficients that are already known from the previous analyses. Explicitly, we find 
\begin{align}
\label{eq:F3:coefficients}
    F^{(3)}(\gamma) \simeq \bigg( 30 &h_{60} F^{(0)}(\gamma) + \frac{3 j_{60}}{20} - \frac{(css)_{33}}{180} + \frac{(jcs)_{42}}{720} - \frac{(jjs)_{51}}{90} + h_{20} \bigg[ \frac{17 j_{40}}{45} - \frac{11 (cs)_{22}}{180} \bigg] \nonumber \\
    + \, &h_{40} \bigg[- \frac{266}{15} h_{20} F^{(0)}(\gamma) + \frac{11 j_{20}}{3}\bigg] 
     + h_{20}^2 \bigg[ -\frac{643 j_{20}}{180} + \frac{37 s_{11}}{15} \bigg] \bigg) L^6 \,. 
\end{align}
In total there are thus five genuine three-loop coefficients at this order, for which we have chosen the set $\{h_{60}, j_{60}, (css)_{33}, (jcs)_{42},  (jjs)_{51}\}$. We now discuss each of these entries in turn:

\begin{itemize}
\item 
The purely hard contribution results again from exponentiation, i.e.~$h_{60} = (h_{20})^3/6$.
\item 
Similar to the analysis at NNLO, the coefficient $j_{60}$ is free from any regulator ambiguities. Its explicit calculation using sophisticated multi-loop techniques will be described in Sec.~\ref{sec:FOcalculation}.
\item 
The $1/(\alpha^3\varepsilon^3)$ singularity in the mixed collinear-soft-soft region, $(css)_{33}$, can be inferred from the NNLO analysis in the previous section in the following way. First, we note that the two-loop coefficient $s_{22}$ is constrained to be $s_{22} = - (cs)_{22}/2$ from pole cancellation at NNLO, with $(cs)_{22}$ given in~\eqref{eq:cs22}. The fact that the soft inverse moments~\eqref{eq:NLO:softmoments} entering~\eqref{eq:purelysoftregion} do not depend on powers of the mass ratio $\bar{u}_0$, then allows one to reconstruct the leading $1/(\alpha^2\varepsilon^2)$ poles of the two endpoint-divergent inverse moments of the $B_c$-meson LCDA individually, 
\begin{align}
    \int_0^\infty \! d \omega \, \frac{m_2^2}{\omega^2} \, \phi_{B,\text{reg}}^{+,(2)}(\omega) &\simeq \frac{2 C_F^2}{\alpha^2\varepsilon^2} \left(\frac{\mu^2}{m_2^2} \right)^{2\varepsilon} \left( \frac{\nu}{m_2} \right)^{2\alpha} + \ldots \,, \nonumber \\
    \int_0^\infty \! d \omega \, \frac{m_2}{\omega} \, \phi_{B,\text{reg}}^{-,(2)}(\omega) &\simeq \frac{6 C_F^2 - C_A C_F}{\alpha^2\varepsilon^2} \left(\frac{\mu^2}{m_2^2} \right)^{2\varepsilon} \left( \frac{\nu}{m_2} \right)^{2\alpha} + \ldots \,. 
     \label{eq:NNLO:softmoments}
\end{align}
These expressions agree with an explicit diagrammatic calculation given in (D.24) of~\cite{Boer:2018mgl}. Notice that in contrast to~\eqref{eq:NLO:softmoments}, we focus here on the leading singularities in the rapidity regulator (as indicated by the dots). The NLO collinear moments are, moreover, also known from~\eqref{eq:NLO:collinearmoments}. Inserting both in the tree-level factorization formula~\eqref{eq:heavytolightFF_naivefacformula_treeHC} then yields the coefficient 
\begin{equation}
    (css)_{33} = \xi_0 C_F^2 \bigg(-20C_F \frac{1+\bar{u}_0}{\bar{u}_0^3} +2 C_A \, \frac{3+2\bar{u}_0}{\bar{u}_0^3} \bigg) \,. 
\end{equation}
\item 
The $1/(\alpha^2 \varepsilon^4)$ singularity of the mixed (hard-collinear)-collinear-soft contribution turns out to vanish, $(jcs)_{42} = 0$. To verify this, we first note that the NNLO analysis implies $(js)_{22} = (js)_{31} = 0$, i.e.~the convolution of the one-loop $B_c$-meson LCDA with the NLO jet functions is completely free of $1/\alpha$ rapidity singularities at the double-logarithmic level. To show that $(jcs)_{42} = 0$, it then suffices to argue that the additional convolution with the NLO collinear LCDA can produce at most a single $1/\alpha$ pole.

For the coefficient discussed in the next item, it is useful to get a deeper understanding why this contribution vanishes from the factorization point-of-view. It is immediately clear for all jet functions $D_i^{(1)}$ with $i \neq 4$ from the fact that the hard-collinear scale is $\mu_{j}^2 \sim 2 \omega E_\eta$. The NLO jet functions are therefore accompanied by a factor $(2\omega E_\eta)^{-\varepsilon}$ on dimensional grounds, as can be verified explicitly in~\eqref{eq:D1NLO} and~\eqref{eq:DmNLO}, which regulates the endpoint divergences in convolutions with the NLO soft functions with the dimensional regulator. This argument cannot be used, however, for the jet function $D_4^{(1)}(u,\omega,\xi)$ due to its dependence on two soft light-cone momenta $\omega$ and $\xi$. Still, the consistency requirements at NNLO demand that some mechanism must be at play to guarantee that also the convolution with the three-particle LCDA $\phi_{3B}^{(1)}(\omega,\xi)$ is free of $1/\alpha$ poles in the double-logarithmic approximation. Restricting to the Abelian limit, we will elaborate on this mechanism in greater detail in App.~\ref{app:D4_omegatozero}. There we use refactorization techniques to show that the terms in $D_4^{(1)}(u,\omega,\xi)$ that are not protected by a factor $\omega^{-\varepsilon}$ in the limit $\omega \to 0$ are only single-logarithmic, i.e. do not contain a double pole in $\varepsilon$.
\item 
Lastly, by the same mechanism as for the one-loop jet function discussed in the previous point, also the coefficient $(jjs)_{51}=0$, i.e.~the NNLO jet functions convoluted with the NLO $B_c$-meson LCDA do not produce a singularity in the rapidity regulator at the double-logarithmic level. 
We note that similar features for another subleading-power NLO jet function were observed in~\cite{Beneke:2019oqx} (see the discussion in Sec.~4.2.3.), and analogous conclusions about its higher-order structure were drawn therein.

\end{itemize}
In summary, we identified a single three-loop coefficient that is currently unknown, $j_{60}$, and needs to be computed to reconstruct the double logarithms of the soft-overlap form factor at N$^3$LO in the SCET approach.

\section{Computation of hard-collinear coefficients}
\label{sec:FOcalculation}

In the previous section we argued that the SCET analysis can provide important cross checks of the integral equations~\eqref{eq:finalintegralequations}, which we derived in~\cite{Bell:2024bxg} with independent methods. Specifically, we found that there are only two unknowns -- the coefficients of the leading singularity in the purely hard-collinear region at two loop ($j_{40}$) and three loop ($j_{60}$) --  that need to be determined to reconstruct the double logarithms up to N$^3$LO in SCET. The computation of these coefficients is by itself rather involved, and we briefly describe the setup we used for this calculation and quote the results in this section.

\subsection{Computational setup}

\begin{figure}[t]
    \centering
    \includegraphics[width=0.32\textwidth]{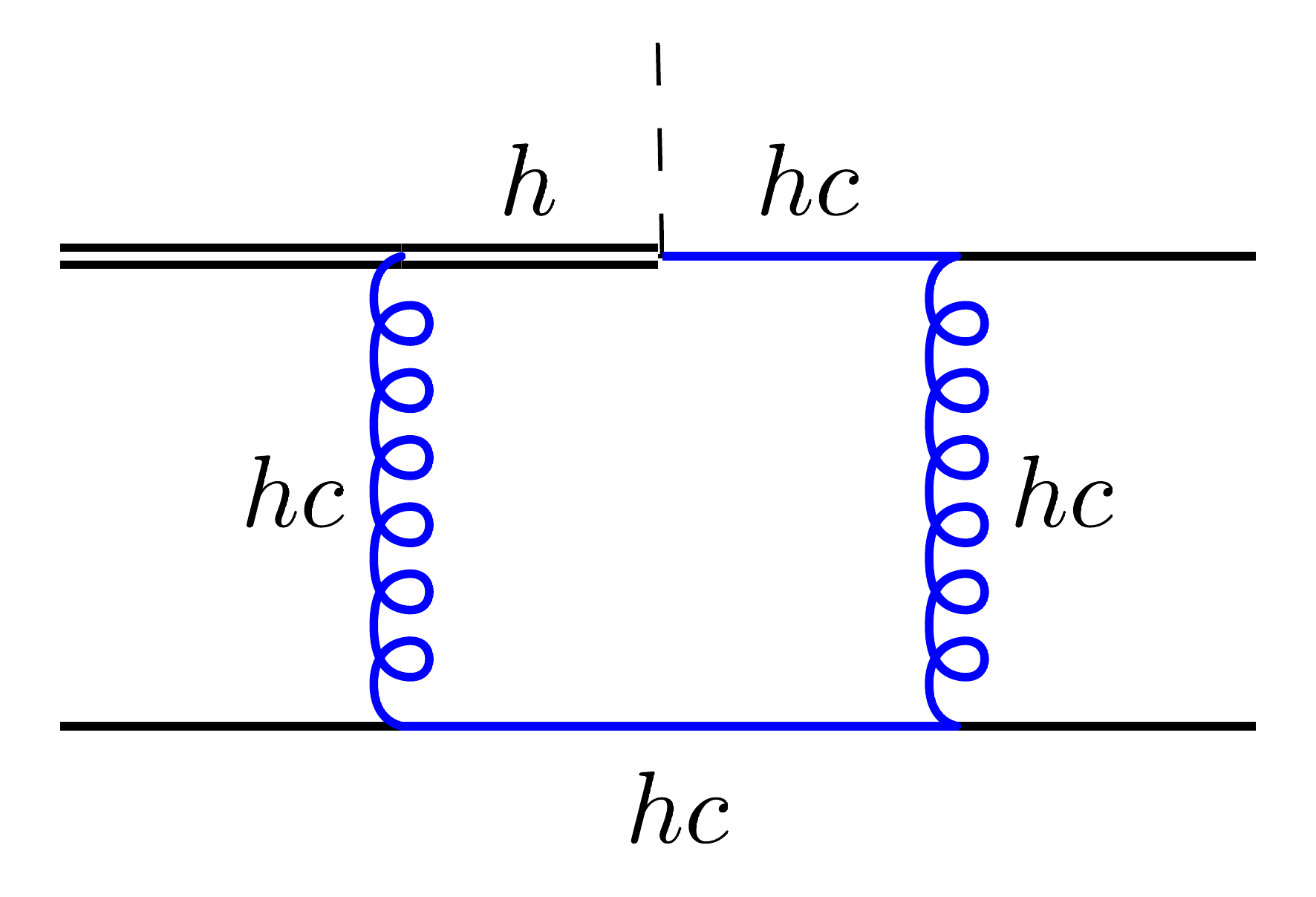}
	\caption{Sample one-loop diagram in the hard-collinear region. The double line represents the heavy $b$-quark.}
	\label{fig:momentumrouting} 
\end{figure}

The details of the partonic calculation, which includes the momentum assignment, the kinematics and the various projectors used to extract the soft-overlap form factor, were already given in Sec.~2~of~\cite{Bell:2024bxg}. Starting from the QCD diagrams, we expand the amplitude in the purely hard-collinear region, i.e.~all loop momenta satisfy the scaling $(k_-,k_\perp,k_+)\sim(1,\lambda,\lambda^2)m_b$ with $\lambda=(m_\eta/m_b)^{1/2}\ll 1$, where  $k_- = \bar n \cdot k$, $k_+ = n \cdot k$ and $k_\perp^\mu$ is transverse to both light-cone directions $n^\mu$ and $\bar{n}^\mu$. The fact that we consider the purely hard-collinear region has important implications: First, the calculation can be performed setting the rapidity regulator to zero from the beginning. Moreover, since the momentum regions are not invariant under shifts of the loop momenta in the presence of heavy-quark propagators (see Fig.~\ref{fig:momentumrouting}), one has to choose a specific momentum routing with the momentum $p_b =m_b v^\mu$ flowing exclusively through $b$-quark propagators. If this is not already the case in the automated diagram-generation process, one has to shift the loop momenta accordingly. Finally, after having performed the power expansion, one can safely set the masses $m_b$ and $m_\eta$ as well as the boost factor $\gamma$ to unity, since the desired hard-collinear coefficients only depend non-trivially on the quark mass ratio $\bar{u}_0$. We also note that the calculation is performed in Feynman gauge.

In the purely hard-collinear region, the light-quark masses can furthermore be neglected in the denominators of the propagators, but not in their numerators because of the contributions from the effective operator $\mathcal{O}_m$. The calculation therefore essentially amounts to a massless five-point function in a special kinematical configuration. We employ a fully automated computational setup consisting of \textsc{QGRAF}~\cite{Nogueira:2006pq}, \textsc{FORM}~\cite{Vermaseren:2000nd,Kuipers:2012rf}, \textsc{FeynCalc}~\cite{Mertig:1990an,Shtabovenko:2016sxi,Shtabovenko:2020gxv,Shtabovenko:2021hjx,Shtabovenko:2023idz,
Shtabovenko:2025lxq} and \textsc{FeynHelpers}~\cite{Shtabovenko:2016whf} for this purpose, and we make use of the in-house \textsc{LoopScalla}~\cite{url:LoopScalla} framework to connect different tools to each other.

In the first step, we generate all contributing diagrams up to three loops with \textsc{QGRAF}, which we may visualize with \textsc{Graphviz}~\cite{url:Graphviz} and \textsc{TikZ-Feynman}~\cite{
Ellis:2016jkw}. The Feynman rules are then inserted with the aid of \textsc{FORM}, which we also use for the evaluation of the Dirac algebra and the power expansion in the hard-collinear region. The color algebra is handled with \textsc{color.h}~\cite{vanRitbergen:1998pn}.

We then extract the propagator denominators present in each diagram and proceed with the topology identification. Specifically, we make use of the new multiloop functionality introduced in \textsc{FeynCalc} 10 that allows one to derive mappings between equivalent topologies, perform partial-fraction decomposition and augment incomplete propagator bases in a straight\-forward fashion. The results of this step are then exported as \textsc{FORM} \texttt{id}-statements and used to convert each amplitude into a sum of scalar loop integrals.

Given the list of final topologies and the corresponding integrals, \textsc{FeynHelpers} can automatically generate run cards for the integration-by-parts (IBP) reduction using \textsc{FIRE}~\cite{Smirnov:2014hma,Smirnov:2019qkx,Smirnov:2023yhb} and \textsc{LiteRed}~\cite{
Lee:2013mka} or \textsc{KIRA}~\cite{Maierhofer:2017gsa,Klappert:2020nbg,Lange:2025fba}. We used \textsc{FIRE} 6.5 with \textsc{FLINT}~\cite{url:Flint} for almost all integral families, except for some three-loop topologies, which were reduced with \textsc{KIRA} instead. The reduction tables are then processed with \textsc{FeynHelpers} and \textsc{FeynCalc}, and we also identify one-to-one mappings between master integrals from different integral families. Furthermore, the reduction rules are expanded in $\varepsilon$ to reduce their size.

Using \textsc{pySecDec}~\cite{Borowka:2017idc,Borowka:2018goh,Heinrich:2021dbf,Heinrich:2023til} we can determine the leading power of the $\varepsilon$-pole in each of the master integrals. By making use of this information during the insertion of the reduction tables, we can significantly reduce the size of the amplitude and the number of contributing master integrals.  In the two-loop case, the IBP reduction proceeds without introducing any spurious poles. This means that it is sufficient to evaluate only the coefficient of the highest pole of each contributing master integral. By adopting a rational-function ansatz, it is then straightforward to reconstruct the $\bar{u}_0$-dependence of each master integral analytically. All that is needed for this purpose is a numerical evaluation of the master integrals at various $\bar{u}_0$-values with sufficient precision, which can be achieved with the aid of \textsc{pySecDec}.

At three loops the situation becomes more involved due to the presence of spurious poles. Knowing that the final result for the amplitude should start at $1/\varepsilon^6$, we encounter spurious poles as high as $1/\varepsilon^{11}$ in intermediate steps of the calculation, which result from an interplay of the chosen basis of master integrals and inverse powers of $\varepsilon$ present in their prefactors. On the one hand, the cancellation of these spurious poles provides a non-trivial check of the consistency of the calculation, but on the other hand it significantly complicates the numerical evaluation of the master integrals, since several coefficients in their $\varepsilon$-expansion need to be evaluated with high precision. Moreover, for the coefficients of the subleading poles, the rational-function ansatz is no longer applicable, since they may contain e.g.~logarithms involving $\bar{u}_0$ and more complicated functions. Even though we expect the final result to be a rational function in $\bar{u}_0$, there is a large amount of cancellations happening in intermediate steps of the calculation.

We therefore refrain from obtaining analytic results for the leading poles of all master integrals, and we rather rely on numerical checks at several $\bar{u}_0$-values to ensure that all spurious poles cancel and the final result for the $1/\varepsilon^6$-pole takes the expected form. Specifically, we find that the spurious poles from $1/\varepsilon^{11}$ up to $1/\varepsilon^{9}$ cancel analytically, the coefficient of the $1/\varepsilon^{8}$ pole is of $\mathcal{O}(10^{-10})$, while for $1/\varepsilon^{7}$ the values vary between $\mathcal{O}(10^{-2})$ and $\mathcal{O}(10^{-5})$ depending on the value of $\bar{u}_0$ being used. We also observe that some particular $\bar{u}_0$-values lead to very large numerical uncertainties for some of the color structures. This can be traced to the symbolic form of these color factors before inserting the numerical values, which usually involve linear combinations of several hundreds of master integrals multiplying $\bar{u}_0$ raised to some very high powers such as 16. For these cases the numerical cancellations become so dramatic, that the precision achievable with \textsc{pySecDec} within a reasonable time frame is not sufficient to clearly see that these terms cancel.

\subsection{Results}

We first test this setup by rederiving the known one-loop hard-collinear coefficient $j_{20}$. At this order,  we start from 31 Feynman diagrams that incorporate 18 integral families. After IBP reduction, the final result can then be reconstructed from the leading poles of three one-loop master integrals.  This yields
\begin{equation}
    j_{20} = \xi_0 \bigg( C_F \frac{10 + 8 \bar{u}_0}{\bar{u}_0^3} - \frac{C_A}{\bar{u}_0^3} \bigg) \, ,
\end{equation}
in agreement with \eqref{eq:NLOhcpoles}.

In the two-loop case, the number of diagrams increases to 722 and we need to handle 374 integral topologies. This leads to 46 contributing master integrals with up to seven propagators. Some of these integrals are simple enough to be calculated directly, but the majority requires a more sophisticated approach. Fortunately, only their leading poles enter the final result and those can be evaluated with \textsc{pySecDec} to a sufficiently high precision. To reconstruct the analytic results, we make an ansatz  in form of a rational function of $\bar{u}_0$, with the allowed denominators being $\bar{u}_0$ or $1-\bar{u}_0$. We then evaluate each integral at multiple carefully chosen points and convert the numerical  results into rational numbers. From here we obtain a system of linear equations for the coefficients of our ansatz. Solving this system allows us to reconstruct the analytic result for the leading pole of each master integral separately. Additional evaluations of the integrals at different  values of $\bar{u}_0$ can furthermore be used as a cross-check. A more detailed description of this procedure can be found in Ref.~\cite{Shtabovenko:2024aum}. Consequently, the two-loop hard-collinear coefficient is found to be
\begin{equation}
    j_{40} = \xi_0 C_F \bigg( C_F \frac{17+15\bar{u}_0}{\bar{u}_0^3} - C_A \frac{5 + \bar{u}_0}{2 \bar{u}_0^3} \bigg) \, , 
\end{equation}
which, when inserted into \eqref{eq:F2:coefficients} together with the remaining coefficients discussed above, reproduces the predicted two-loop expression for the soft-overlap form factor given in the second line of \eqref{eq:FO}.

Finally, at three loops there are $20,\!759$ diagrams to evaluate that give rise to $6,\!276$ integral families. We find $1,\!671$ three-loop master integrals that contribute to the final result. Furthermore, there are two complications that did not arise at lower loop level. First of all, many integrals contribute with their subleading poles as discussed earlier. Even though one expects that the final result may only contain rational functions of $\bar{u}_0$, this does not automatically apply to intermediate results. The cancellations of non-rational contributions are difficult to check numerically, because they require very high precision. In total, we find that $1,\!202$ integrals contribute with their leading pole, 399 integrals with their first two coefficients, 50 with three, 16 with four and 4 integrals with five coefficients. Second, the numerical evaluation of the integrals using \textsc{pySecDec} -- especially for cases with 12 propagators and dots -- is much more challenging as compared to the two-loop computation. For some integrals and $\bar{u}_0$-values, it is hardly possible to go beyond a relative precision of $10^{-5}$. The rational functions of $\bar{u}_0$ multiplying such integrals often turn out to be very complicated, which leads to large numerical cancellations.

\begin{figure}[t]
				\centering
                \includegraphics[width=0.6\textwidth]{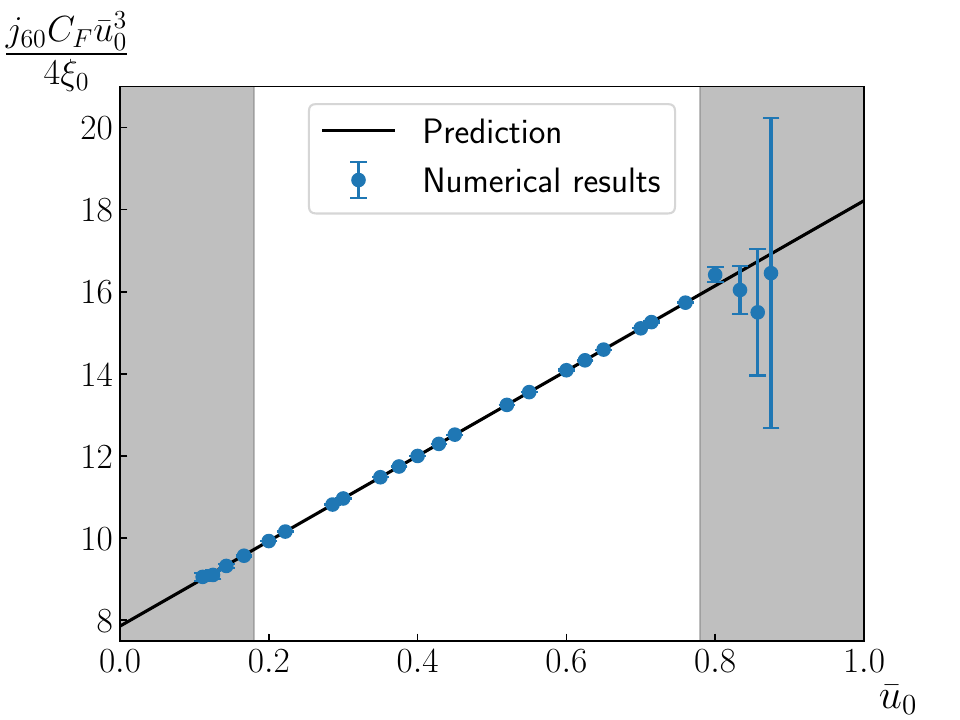}
	\caption{Result for the three-loop hard-collinear coefficient $j_{60}$ as a function of the quark-mass ratio $\bar{u}_0$. The dots show the numerical results of our three-loop computation, and the line displays the expression~\eqref{eq:j60:expectation} that is consistent with the three-loop prediction of the integral equations. 
    }
	\label{fig:j60fit} 
\end{figure}

In view of these complications, we chose to evaluate the three-loop hard-collinear coefficient numerically for 25 values of $\bar{u}_0$ and to reconstruct its analytic $\bar{u}_0$-dependence by fitting these results. To quantify the uncertainties of the numerical results, we combined the Monte-Carlo integration errors from \textsc{pySecDec} for the individual master integrals in quadrature. The result for the ratio $j_{60}C_F\bar{u}_0^3/4\xi_0$ is shown in Fig.~\ref{fig:j60fit}, where the data points clearly exhibit a simple linear behavior. By performing a fit to the central 17 points, as indicated in the figure, and using the sidebands to further validate the assumed linear dependence, we obtain
\begin{align}
\label{eq:j60:fit}
     \frac{j_{60}C_F\bar{u}_0^3}{4\xi_0} = (7.856 \pm 0.005) + (10.366 \pm 0.013) \bar{u}_0\,. 
\end{align}
This result can be compared with the one obtained from the three-loop prediction of the integral equations given in the third line of \eqref{eq:FO}, upon inserting all relevant coefficients into \eqref{eq:F3:coefficients} and solving this relation for $j_{60}$. This yields
\begin{align}
\label{eq:j60:expectation}
     \frac{j_{60}C_F\bar{u}_0^3}{4\xi_0} = C_F^3  \left( C_F\frac{ 37 + 34 \bar{u}_0}{9}-  C_A\frac{ 13 +4\bar{u}_0}{18} \right) = 7.857 + 10.359 \, \bar{u}_0 \,. 
\end{align}
We take this as a strong indication that the SCET analysis correctly reproduces the prediction of the integral equations at this highly non-trivial perturbative order. Further details of the fit for individual color structures of the three-loop coefficient can be found in App.~\ref{app:j60}.

\section{Discussion}
\label{sec:discussion}

Having established that the SCET analysis reproduces the double logarithms of the soft-overlap form factor up to three-loop order, one may ask how the specific structure in \eqref{eq:Ffinaldecomp}, involving the two auxiliary functions $f_{1,2}$, can be interpreted within SCET. In this section, we show that these functions can be related to endpoint-divergent inverse moments of the $B_c$-meson LCDA\footnote{As was already observed in~\cite{Bell:2024bxg}, a similar analysis could also be formulated for $\eta_c$-meson LCDA.}, provided that the convolution integrals are regularized by an appropriate cutoff $\kappa$. More specifically, we demonstrate that the complete information on the double-logarithmic corrections is encoded in a specific set of evolution equations involving derivatives with respect to both the factorization scale $\mu$ and the cutoff $\kappa$.

We begin by rewriting the integral equations~\eqref{eq:finalintegralequations} in an equivalent differential form. In the notation of~\cite{Bell:2024bxg}, they read
\begin{align}
\label{eq:finalintegralequations:differentialform}
    \left( \partial_\rho \partial_\eta + \eta \partial_\eta - 1 \right) g_1(\rho,\eta) &= 0 \,, \nonumber\\
    \left( \partial_\rho \partial_\eta + \eta \partial_\eta - 1 \right) \left( g_2(\rho,\eta) +\frac12 \right) &=
     - \frac{1}{ 4N_c C_F} \left( g_1(\rho,\eta) - 1 \right) \,, 
\end{align}
where
\begin{equation}
\rho = \sqrt{\frac{\alpha_s C_F}{2\pi}} \, \ln \left(\frac{q_+ p_-}{m_2^2}\right) \,, \qquad 
 \eta = \sqrt{\frac{\alpha_s C_F}{2\pi}} \,\ln \left(\frac{p_-}{q_-}\right)  \,, 
\end{equation}
are logarithmic variables in the light-cone components $q_+$ and $q_-$, and we introduced new symbols for the auxiliary functions with
\begin{equation}
 f_1(q_+,q_-) = g_1(\rho,\eta) \,, \qquad f_2(q_+,q_-) = g_2(\rho,\eta) \,. 
 \end{equation}
The information on the integration limits in~\eqref{eq:finalintegralequations} is then encoded in the boundary conditions
\begin{align}
\label{eq:PDEs_boundaries}
    g_1(\rho,\eta=0) &= 1 \,, & \partial_\eta g_1(\rho,\eta)\big\vert_{\rho = \eta} = 0 \,, \nonumber \\
    g_2(\rho,\eta=0) &= 1 \,, & \partial_\eta g_2(\rho,\eta)\big\vert_{\rho = \eta} = 0 \,. 
\end{align}
Notice also that one can identify $p_- \approx p_{\eta_-} \approx p_{2-}$ in the double-logarithmic approximation.

The starting point of our analysis is Eq.~\eqref{eq:purelysoftregion}, which we repeat here for convenience,
\begin{align}
\label{eq:softcontribution:repetition}
  \frac{F(\gamma)\big\vert_{\text{soft}}}{\xi_0}
  \simeq &\int_0^\infty \!\! d \omega \, \frac{m_2}{\omega} \Bigg\{ \!\frac{1+\bar{u}_0}{\bar{u}_0^3} \bigg(\phi_B^-(\omega)
  + \frac{m_2}{\omega} \phi_B^+(\omega) \bigg) 
  - \frac{\phi_B^+(\omega)}{\bar{u}_0^2} 
  + \frac{C_A}{2C_F \bar{u}_0^3} \, \phi_B^+(\omega) \, \bigg( 1 - \frac{m_2}{\omega} \!\bigg) \!\Bigg\} \;. 
\end{align}
This identity expresses the all-order purely soft contribution to the form factor $F(\gamma)$ in terms of (finite as well as endpoint-divergent) inverse moments of the two-particle $B_c$-meson LCDA $\phi_B^\pm(\omega)$. It is valid in the double-logarithmic approximation as indicated by the $\simeq$ symbol. The key observation is that the right-hand side of~\eqref{eq:softcontribution:repetition} takes a similar form as~\eqref{eq:Ffinaldecomp} if we identify the endpoint-divergent moments in this relation with the functions $f_{1,2}(m_2,m_2)$ that appear in~\eqref{eq:Ffinaldecomp} according to
\begin{align}
\label{eq:f12replacement}
    \int_0^\infty \! d \omega \, \frac{m_2^2}{\omega^2} \, \phi_B^+(\omega) &\longrightarrow \exp\bigg\{ -\frac{\alpha_s C_F}{4\pi} L^2\bigg\} \times f_1(m_2,m_2) \;, \nonumber\\
    \int_0^\infty \! d \omega \, \frac{m_2}{\omega} \bigg(\phi_B^-(\omega)
  + \frac{m_2}{\omega} \phi_B^+(\omega) \bigg) &\longrightarrow \exp\bigg\{ -\frac{\alpha_s C_F}{4\pi} L^2\bigg\} \times 2 f_2(m_2,m_2) \,, 
\end{align}
and the endpoint-finite first inverse moment of $\phi_B^+(\omega)$ with 
\begin{equation}
\label{eq:lamBreplacement}
    \int_0^\infty \! d \omega \, \frac{m_2}{\omega} \, \phi_B^+(\omega) \longrightarrow \exp\bigg\{ -\frac{\alpha_s C_F}{4\pi} L^2\bigg\} \,.
\end{equation}
While this identification may seem adhoc at first sight, we will motivate it in the following by observing that the functions $f_{1,2}$ mimic the logarithmic structure of inverse moments supplemented with a cutoff $\kappa$ to regularize the endpoint divergences. The differential equations~\eqref{eq:finalintegralequations:differentialform} are then equivalent to combined evolution equations in the cutoff $\kappa \ll m_2$ and the usual renormalization scale $\mu$ of the LCDA $\phi_B^\pm(\omega;\mu)$.

Starting with the simpler case, the RG equation of the leading-twist LCDA $\phi_B^+(\omega;\mu)$ reads
\begin{align}
    \frac{d}{d\ln \mu} \,\phi_B^+(\omega;\mu) = - \int_0^\infty \! d\omega'~\gamma_+(\omega,\omega';\mu)\,\phi_B^+(\omega';\mu) \;, 
    \label{eq:phiplus:evolution}
\end{align}
with the one-loop anomalous dimension~\cite{Lange:2003ff} 
\begin{align}
    \gamma_+^{(1)}(\omega,\omega';\mu) = \bigg[\Gamma_{\rm cusp}^{(1)} \ln \frac{\mu}{\omega} + \gamma^{(1)} \bigg] \delta(\omega-\omega') - \Gamma_{\rm cusp}^{(1)} \, \omega \,\Gamma(\omega,\omega') \; .
\end{align}
Here, $\Gamma_{\rm cusp}^{(1)} = 4C_F$ is the one-loop cusp anomalous dimension in the fundamental representation, and the kernel $\Gamma(\omega,\omega')$ is given by
\begin{align}
    \Gamma(\omega,\omega') &= \bigg[ \frac{\theta(\omega-\omega')}{\omega(\omega-\omega')} + \frac{\theta(\omega'-\omega)}{\omega'(\omega'-\omega)} \bigg]_+ \;, 
\end{align}
where the plus-distribution acts on a test function $f$ according to $f(\omega') \to f(\omega') - f(\omega)$. 

Focusing on the double-logarithmic contributions, we can simplify the RG equation as follows. First, we keep the usual logarithmically-enhanced contribution $\sim \ln \mu$ in the anomalous dimension $\gamma_+(\omega,\omega';\mu)$. In addition, since we consider endpoint-sensitive moments and the cutoff $\kappa \ll m_2$ is much smaller than the typical soft scale of the emissions, we also keep terms that generate logarithms $\sim \ln \omega$ from the integration over $\omega'$. The resulting asymptotic evolution equation reads 
in this approximation
\begin{equation}
       \frac{d}{d\ln \mu} \,\phi_B^+(\omega;\mu) \simeq -\frac{\alpha_s}{4 \pi}\,\Gamma_{\rm cusp}^{(1)} \bigg(\ln \frac{\mu}{\omega} \, \phi_B^+(\omega;\mu) - \omega \int_\omega^\infty \! \frac{d\omega'}{\omega'^2} \, \phi_B^+(\omega';\mu) \bigg) \;. 
    \end{equation} 
Here, it is important that $\phi_B^+(\omega;\mu) \sim \omega$ for small $\omega$, such that the integral in the last term generates a logarithmic sensitivity to $\omega$. 

Next, we rescale
\begin{equation}
\label{eq:Sudakov:rescaling}
        \phi_B^+(\omega;\mu) \equiv \exp \left\{ - \frac{\alpha_s C_F}{2\pi} \ln^2 \frac{\mu}{\mu_0} \right\} \tilde{\phi}_B^+(\omega;\mu,\mu_0) \;,
    \end{equation}
with some fixed reference scale $\mu_0 \sim m_2$, and from now on we insert the explicit expression for the one-loop cusp anomalous dimension for convenience. This factors out the universal exponentiated soft-gluon corrections, which combine with the hard and hard-collinear counterparts to generate the global Sudakov exponent in~\eqref{eq:Ffinaldecomp}. In other words, the remaining double logarithms in the function $\tilde{\phi}_B^+(\omega;\mu,\mu_0)$ can be related to the soft-quark dynamics and its non-trivial interplay with soft-gluon emissions. 
For the endpoint-finite first inverse moment, in particular, only the standard Sudakov factor contributes to the double logarithms, which explains the identification in~\eqref{eq:lamBreplacement}. The evolution equation that governs double logarithms associated with non-trivial soft-quark exchanges then becomes 
\begin{equation}
       \frac{d}{d\ln \mu} \, \tilde{\phi}_B^+(\omega;\mu,\mu_0) \simeq \frac{\alpha_s C_F}{\pi} \bigg( \ln \frac{\omega}{\mu_0} \, \tilde{\phi}_B^+(\omega;\mu,\mu_0) + \omega \int_\omega^\infty \! \frac{d\omega'}{\omega'^2} \,\tilde{\phi}_B^+(\omega';\mu,\mu_0) \bigg) \,, 
\end{equation}
and we stress that $\ln\omega/\mu_0$ is considered to be a large logarithm here, because the moments we consider are sensitive to the endpoint region with $\omega\ll\mu_0\sim m_2$. Lastly, we define the second inverse moment of $\tilde{\phi}_B^+(\omega;\mu,\mu_0)$ with a lower cutoff $\kappa \ll m_2$,
\begin{equation}
        g^+(\kappa;\mu,\mu_0) \equiv \int_\kappa^\infty \! \frac{d\omega}{\omega^2} \, \tilde{\phi}_B^+(\omega;\mu,\mu_0) \,, 
\end{equation}
which fulfills the differential equation
\begin{equation}
        \bigg( \frac{d^2}{d\ln \kappa \, d \ln \mu} + \frac{\alpha_s C_F}{\pi} \ln \frac{\mu_0}{\kappa} \frac{d}{d\ln\kappa} + \frac{\alpha_s C_F}{\pi} \bigg) \, g^+(\kappa;\mu,\mu_0) = 0 \;. 
\end{equation}
Upon identifying $\rho = \sqrt{\frac{\alpha_s C_F}{2\pi}} \ln\frac{\mu^2}{\mu_0^2}$ and $\eta = \sqrt{\frac{\alpha_s C_F}{2\pi}} \ln\frac{\mu_0}{\kappa}$, this second order differential equation in the virtuality $\mu$ and the longitudinal cutoff $\kappa$ reproduces the differential equation for $g_1(\rho,\eta)$ in~\eqref{eq:finalintegralequations:differentialform}, which motivates the replacement in the first line of~\eqref{eq:f12replacement}. 
The first boundary condition in Eq.~\eqref{eq:PDEs_boundaries} then translates to $g^+(\kappa = \mu_0;\mu,\mu_0) = 1/m_2^2$, i.e. the inverse moment reduces to its LO expression if the cutoff is \emph{not} parametrically smaller than the soft scale. The second boundary condition implies that all radiative corrections to the LCDA $\tilde{\phi}_B^+(\omega;\mu,\mu_0)$ vanish at the specific point $\mu^2/\mu_0^2 = \mu_0/\omega$.

In a similar way, one can show that the RG equation for the subleading LCDA $\phi_B^-(\omega;\mu)$ then leads to a differential equation for its cut-off inverse moment that is equivalent to the integral equation for the function $f_2$ in~\eqref{eq:finalintegralequations}. The key difference to the previous case is, that $\phi_B^-(\omega;\mu)$ is not diagonal under RG evolution. The one-loop anomalous dimension contains mixing terms with the leading LCDA $\phi_B^+(\omega;\mu)$ (for non-vanishing quark masses) and the three-particle LCDA $\phi_{3B}(\omega,\xi;\mu)$, 
\begin{align}
        \frac{d}{d\ln\mu} \phi_B^-(\omega;\mu) = - &\int_0^\infty \! d\omega' \, \gamma_-(\omega,\omega';\mu) \phi_B^-(\omega';\mu) \nonumber \\
        - &\int_0^\infty \! d\omega' \, \gamma_{-+}(\omega,\omega';\mu) \phi_B^+(\omega';\mu) \nonumber \\ 
        - &\int_0^\infty \! d\omega' d\xi' \, \gamma_{-3}(\omega,\omega',\xi';\mu) \phi_{3B}(\omega',\xi';\mu)  \;. 
    \end{align}
Here, the first two anomalous dimensions read~\cite{Descotes-Genon:2009jif,Bell:2008er}
\begin{align}
        \gamma_-^{(1)}(\omega,\omega';\mu) &= \gamma_+^{(1)}(\omega,\omega';\mu) - 4C_F \frac{\theta(\omega'-\omega)}{\omega'} \; \nonumber \\
        \gamma_{-+}^{(1)}(\omega,\omega';\mu) &= -4C_F \bigg[ \frac{m_2 \theta(\omega'-\omega)}{\omega'^2} \bigg]_+ \;, 
    \end{align}
and the mixing contribution with $\phi_{3B}(\omega,\xi;\mu)$ is~\cite{Descotes-Genon:2009jif}
\begin{align}
        \gamma_{-3}^{(1)}(\omega,\omega',\xi';\mu) = 4 \bigg[ \frac{1}{\omega'} \bigg\{ (C_A - 2 C_F) &\bigg[ \frac{1}{\xi'^2} \frac{\omega - \xi'}{\omega' + \xi' - \omega} \theta(\xi' - \omega) + \frac{\theta(\omega'+\xi'-\omega)}{(\omega' + \xi')^2} \bigg] \\
        -C_A &\bigg[ \frac{\theta(\omega'+\xi'-\omega)}{(\omega' + \xi')^2} - \frac{\theta(\omega - \omega') - \theta(\omega - \omega' - \xi')}{\xi'^2} \bigg] \bigg\} \bigg]_+ \;. \nonumber 
\end{align}
Expanding all contributions for small $\omega$ and keeping only the logarithmically-enhanced contributions to the relevant first inverse moment leads to
\begin{align}
        \frac{d}{d\ln\mu} \phi_B^-(\omega;\mu) \simeq - &\frac{\alpha_s C_F}{\pi} \bigg( \ln \frac{\mu}{\omega} \, \phi_B^-(\omega;\mu) - \int_\omega^\infty \! \frac{d\omega'}{\omega'} \, \phi_B^-(\omega';\mu) \bigg) \\
        + &\frac{\alpha_s C_F}{\pi} m_2  \int_\omega^\infty \! \frac{d\omega'}{\omega'^2} \, \phi_B^+(\omega';\mu) + \frac{\alpha_s C_A}{\pi} \int_\omega^\infty \! \frac{d\omega'}{\omega'} \int_0^\infty \! \frac{d\xi'}{\xi'^2} \, \phi_{3B}(\omega',\xi';\mu) \,. \nonumber 
\end{align}
Recall that $\phi_{3B}(\omega,\xi;\mu)$ does not vanish for $\omega \to 0$ in the non-relativistic setup, as discussed in Sec.~\ref{subsec:HMEs}, and hence the last term cannot be disregarded. 

Similar to~\eqref{eq:softeom}, the equation of motion~\eqref{eq:softeom1} allows us to replace the specific inverse moment of $\phi_{3B}(\omega,\xi;\mu)$ with moments of $\phi_B^+(\omega;\mu)$. Dropping contributions of subleading logarithmic order, we use 
\begin{equation}
        2 \int_\omega^\infty \! \frac{d\omega'}{\omega'} \int_0^\infty \! \frac{d\xi'}{\xi'^2} \, \phi_{3B}(\omega',\xi';\mu) \simeq \int_0^\infty \! \frac{d\omega'}{\omega'} \phi_B^+(\omega';\mu) - m_2 \int_\omega^\infty \! \frac{d\omega'}{\omega'^2} \phi_B^+(\omega';\mu) \;, 
    \end{equation}
to arrive at the double-logarithmic evolution equation
\begin{align}
        \frac{d}{d\ln\mu} \phi_B^-(\omega;\mu) \simeq - &\frac{\alpha_s C_F}{\pi} \bigg( \ln \frac{\mu}{\omega} \, \phi_B^-(\omega;\mu) - \int_\omega^\infty \! \frac{d\omega'}{\omega'} \, \phi_B^-(\omega';\mu) \bigg) \nonumber \\
        + &\frac{\alpha_s}{\pi} \big(C_F - \frac{C_A}{2}\big) m_2  \int_\omega^\infty \! \frac{d\omega'}{\omega'^2} \, \phi_B^+(\omega';\mu) + \frac{\alpha_s}{\pi} \frac{C_A}{2} \int_0^\infty \! \frac{d\omega'}{\omega'} \phi_B^+(\omega';\mu) \,. 
\end{align}
Notice that the lower integration boundary $\omega$ can be set to zero in the endpoint-finite first inverse moment of $\phi_B^+(\omega;\mu)$ as it does not feature a logarithmic enhancement.
Factoring out the soft-gluon contributions similar to~\eqref{eq:Sudakov:rescaling}, we arrive at the evolution equation 
\begin{align}
        \frac{d}{d\ln\mu} \tilde{\phi}_B^-(\omega;\mu,\mu_0) \simeq \, &\frac{\alpha_s C_F}{\pi} \bigg( \ln \frac{\omega}{\mu_0} \, \tilde{\phi}_B^-(\omega;\mu,\mu_0) + \int_\omega^\infty \! \frac{d\omega'}{\omega'} \, \tilde{\phi}_B^-(\omega';\mu,\mu_0) \bigg) \nonumber \\
        + &\frac{\alpha_s}{\pi} \big(C_F - \frac{C_A}{2}\big) m_2  \int_\omega^\infty \! \frac{d\omega'}{\omega'^2} \, \tilde{\phi}_B^+(\omega';\mu,\mu_0) + \frac{\alpha_s}{\pi} \frac{C_A}{2}\,, 
    \end{align}
that captures double-logarithmic contributions from soft-quark exchanges and intertwined soft-gluon corrections.  
The cut-off first inverse moment of the subleading LCDA $\phi_B^-(\omega;\mu)$, 
\begin{equation}
        g^-(\kappa;\mu,\mu_0) \equiv \int_\kappa^\infty \! \frac{d\omega}{\omega} \, \tilde{\phi}_B^-(\omega;\mu,\mu_0) \; , 
\end{equation}
then fulfills
\begin{align}
        &\bigg(\frac{d^2}{d\ln\mu \, d \ln \kappa} - \frac{\alpha_s C_F}{\pi} \ln \frac{\kappa}{\mu_0} \frac{d}{d\ln \kappa} + \frac{\alpha_s C_F}{\pi} \bigg) g^-(\kappa;\mu,\mu_0) \nonumber \\
        = - &\frac{\alpha_s}{\pi} \bigg( \big(C_F - \frac{C_A}{2}\big) m_2 \, g^+(\kappa;\mu,\mu_0) 
        + \frac{C_A}{2} \bigg) \,. 
\end{align}
After adding the contribution from $g^+(\kappa;\mu)$, this gives the differential equation for the function $g_2(\rho,\eta)$ in~\eqref{eq:finalintegralequations:differentialform}, and motivates the replacement in the second line of~\eqref{eq:f12replacement}. Hence, we showed that the non-standard evolution equations~\eqref{eq:finalintegralequations:differentialform} in the longitudinal momenta can be derived from standard RGEs of the $B_c$-meson LCDA, provided their (divergent) inverse moments are regularized by a cutoff.

\section{Conclusion}
\label{sec:conclusion}

We analyzed the structure of large double-logarithmic corrections to the soft-overlap contribution in heavy-to-light transition form factors at large hadronic recoil. As demonstrated in our previous work~\cite{Bell:2024bxg} -- which focused on a perturbative non-relativistic description of $B_c \to \eta_c$ transitions in the scale hierarchy $m_b \gg m_c \gg \Lambda_{\rm QCD}$ -- these double logarithms originate from two distinct yet intertwined sources: exponentiated soft-gluon corrections, giving rise to familiar Sudakov factors, and specific soft-quark configurations associated with rapidity-ordered spectator-quark propagators, which lead to nested integrals over longitudinal light-cone momenta. The resulting double-logarithmic series for the soft-overlap form factor is encoded in a coupled set of integral equations for two auxiliary functions $f_{1,2}$, given in Eqs.~\eqref{eq:Ffinaldecomp} and \eqref{eq:finalintegralequations}.

The present article is a sequel to~\cite{Bell:2024bxg} and pursues two main objectives. First, we performed an independent fixed-order validation of the diagrammatic resummation derived in our previous work up to next-to-next-to-next-to-leading order (N$^3$LO). Second, we reanalyzed the form factor with methods from Soft-Collinear Effective Theory (SCET). Although the resulting bare factorization formula contains endpoint-divergent convolution integrals, we demonstrated that it correctly captures the endpoint dynamics up to three-loop order with appropriate regulators in place. In other words, the breakdown of standard SCET techniques does not appear to indicate the presence of additional momentum modes, but it rather reflects the difficulty of clearly separating the soft and collinear dynamics. Furthermore, we established an intriguing connection between the endpoint-divergent inverse moments of $B_c$-meson light-cone distribution amplitudes and the auxiliary functions $f_{1,2}$.

The fixed-order validation exploits the fact that the singularities in the dimensional regulator $\varepsilon$ and the rapidity regulator $\alpha$ must cancel in the sum over all momentum regions. While we demonstrated this cancellation explicitly at one-loop order, we subsequently used these pole-cancellation arguments to reduce the number of coefficients that are required to reconstruct the double logarithms at two and three loops. By exploiting the underlying factorization structure, we found that only two coefficients need to be determined to extend the SCET analysis to N$^3$LO, which we chose to be the leading $1/\varepsilon^4$ ($1/\varepsilon^6$) singularity of the purely hard-collinear region at two (three) loops. We then computed these coefficients using a set of multi-loop techniques and found complete agreement with the predictions from the integral equations. This constitutes a highly non-trivial check of both our previous diagrammatic analysis~\cite{Bell:2024bxg} and the SCET framework developed in the present work and in~\cite{Boer:2018mgl}.

We furthermore observed that the resummation of double logarithms in the soft-overlap form factor can be derived from a novel set of evolution equations for inverse moments of $B_c$-meson LCDA. More precisely, we showed that -- upon regularizing the endpoint-divergent second inverse moment of $\phi_B^+(\omega;\mu)$ and the first inverse moment of $\phi_B^-(\omega;\mu)$ with a longitudinal momentum cutoff $\kappa$ -- the integral equations for the functions $f_{1,2}$ become equivalent to a coupled set of evolution equations in both the renormalization scale $\mu$ and the cutoff $\kappa$. This identification may provide guidance towards an all-order renormalized factorization theorem for the soft-overlap form factor, which is currently unknown. More generally, it suggests a practical and systematic framework for computing double-logarithmic corrections from soft-quark configurations in a range of power-suppressed processes for which all-order results are presently not available.

\subsubsection*{Acknowledgements}

We thank V.~Magerya for pointing out how to accelerate the procedure of determining the power of the leading $\varepsilon$-pole with \textsc{pySecDec} without evaluating the integral numerically. 
The research of GB, TF, DH and VS was supported by the Deutsche Forschungs\-gemeinschaft (DFG, German Research Foundation) under grant 396021762 - TRR 257. GB, TF and VS were also supported by Germany’s Excellence Strategy – Cluster of Excellence ``Color meets Flavor'', EXC 3107 – Project-ID 533766364. The research of PB was funded by the European Union's Horizon 2020 Research and Innovation Program under the Marie Sk\l{}odowska-Curie grant agreement No.101146976. All Feynman diagrams were produced with \texttt{FeynGame} \cite{Harlander:2020cyh, Harlander:2024qbn, Bundgen:2025utt}.

\appendix

\section{Coefficients of the method-of-regions analysis}
\label{app:coefficients}

In this appendix, we collect all coefficients of the method-of-regions analysis in Sec.~\ref{sec:polecancellation} 
up to next-to-next-to-leading order. Since the hard matching coefficient $H(m_b, E_\eta)$ is multiplicative, see~\eqref{eq:SCET1factorization}, and exponentiates to all orders, see~\eqref{eq:hard_exponentiation}, we only need its one-loop coefficient $h_{20} = -C_F$ in the double-logarithmic approximation, and list the hard-collinear, collinear and soft contributions below.

\subsection{Next-to-leading order}

At next-to-leading order, the leading singularity in $\varepsilon$ of the hard-collinear regions reads
\begin{equation}
    j_{20} = \xi_0 \bigg( 2 C_F \frac{5 + 4 \bar{u}_0}{\bar{u}_0^3} - \frac{C_A}{\bar{u}_0^3} \bigg) \,.
\end{equation}
The collinear and soft low-energy regions develop singularities in the dimensional regulator $\varepsilon$ and the rapidity regulator $\alpha$. The cancellation of the latter relates the coefficients 
\begin{equation}
    s_{11} = - c_{11} = \xi_0 \bigg(6C_F \frac{1+\bar{u}_0}{\bar{u}_0^3} - C_A \, \frac{1}{\bar{u}_0^3} \bigg) \,.
\end{equation}
The regulator prescription in~\eqref{eq:regulator} ensures that all double poles in $\varepsilon$ are part of the soft region, and hence $c_{20} =0$. The respective coefficient of the soft contribution is given by 
\begin{equation}
    s_{20} = \xi_0 \left(- C_F \frac{8+7\bar{u}_0}{\bar{u}_0^3} + \frac{C_A}{\bar{u}_0^3} \right) \,.
\end{equation}

\subsection{Next-to-next-to-leading order}

At next-to-next-to-leading order, the leading singularity in $\varepsilon$ of the hard-collinear region reads
\begin{equation}
    j_{40} = \xi_0 C_F \bigg( C_F \frac{17+15\bar{u}_0}{\bar{u}_0^3} - C_A \frac{5 + \bar{u}_0}{2 \bar{u}_0^3} \bigg) \,.
\end{equation}
The leading poles of the low-energy contributions in the rapidity regulator $\alpha$ are given by
\begin{equation}
    s_{22} = c_{22} = -\frac{(cs)_{22}}{2} = \xi_0 C_F \bigg( 8 C_F \frac{1+\bar{u}_0}{\bar{u}_0^3} - C_A \frac{2 + \bar{u}_0}{\bar{u}_0^3} \bigg) \,,
\end{equation}
whereas their remaining coefficients, subleading in $\alpha$, are not fully determined from pole cancellation. In particular, we find
\begin{equation}
    c_{31} +  \frac{(cs)_{31}}{2} = -s_{31} -  \frac{(cs)_{31}}{2} = \xi_0 C_F \bigg( 18 C_F \frac{1+\bar{u}_0}{\bar{u}_0^3} -  C_A \frac{8+3\bar{u}_0}{2\bar{u}_0^3} \bigg) \,,   
\end{equation}
which satisfy the relation $c_{31} + s_{31} + (cs)_{31} = 0$. The leading poles in $\varepsilon$, on the other hand, fulfill the constraint
\begin{equation}
    c_{40} + s_{40} + (cs)_{40} = \xi_0 C_F \bigg( C_F \frac{100+99\bar{u}_0}{2\bar{u}_0^3} - 7 C_A \frac{3+\bar{u}_0}{2\bar{u}_0^3} \bigg) \,.
\end{equation}
Additionally, there are six coefficients associated with mixed regions that involve hard-collinear contributions. 
We find that both the (hard-collinear)-collinear and (hard-collinear)-soft regions are free of $1/\alpha$ poles, i.e. $(jc)_{22} = (js)_{22} = 0$ and $(jc)_{31} = (js)_{31} = 0$. The remaining two coefficients obey
\begin{equation}
    (jc)_{40} + (js)_{40} = \xi_0 C_F \bigg( -C_F \frac{66+64\bar{u}_0}{\bar{u}_0^3} + C_A \frac{13+4\bar{u}_0}{\bar{u}_0^3} \bigg) \,.
\end{equation}

\section{The \texorpdfstring{$\omega \to 0$}{} limit of the jet function \texorpdfstring{$D_4^{(1)}(u,\omega,\xi)$}{}}
\label{app:D4_omegatozero}

In Sec.~\ref{subsec:N3LO} it was discussed that the cancellation of singularities between the different momentum regions implies the relations $(js)_{22} = (js)_{31} = 0$, i.e.~at NNLO the mixed soft and hard-collinear region is free of poles in the rapidity regulator $\alpha$ at the double-logarithmic level. From the SCET perspective, dimensional analysis dictates that all jet functions $D_i^{(1)}$ with $i \neq 4$ are proportional to a factor $(2\omega E_\eta)^{-\varepsilon}$, which regularizes endpoint divergences from $\omega \to 0$ in convolution integrals with the NLO $B_c$-meson LCDA with the parameter $\varepsilon$. Due to the second light-cone momentum $\xi$ of the soft gluon, the dimensional argument cannot be applied to the convolution of $D_4^{(1)}(u,\omega,\xi)$ 
with the $B_c$-meson LCDA $\phi_{3B}^{(1)}(\omega,\xi)$. However, from the absence of $1/\alpha$ poles in this momentum region, one concludes that also this convolution must be regularized by the dimensional regulator $\varepsilon$ in the double-logarithmic approximation.

\begin{figure}[t]
	\centering
	\includegraphics[trim = 20 100 20 100,clip,width = 0.24\textwidth]{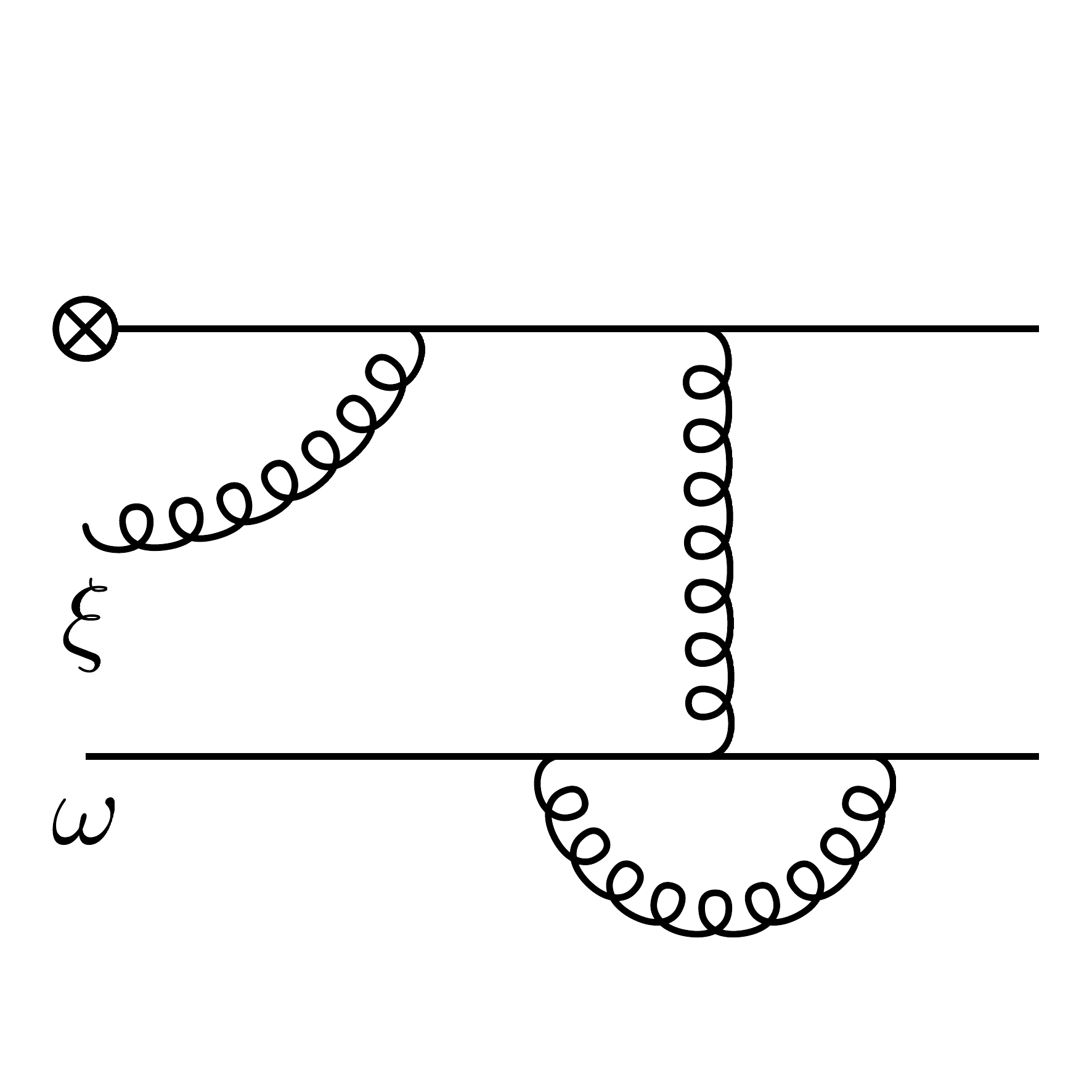}
    \includegraphics[trim = 20 100 20 100,clip,width = 0.24\textwidth]{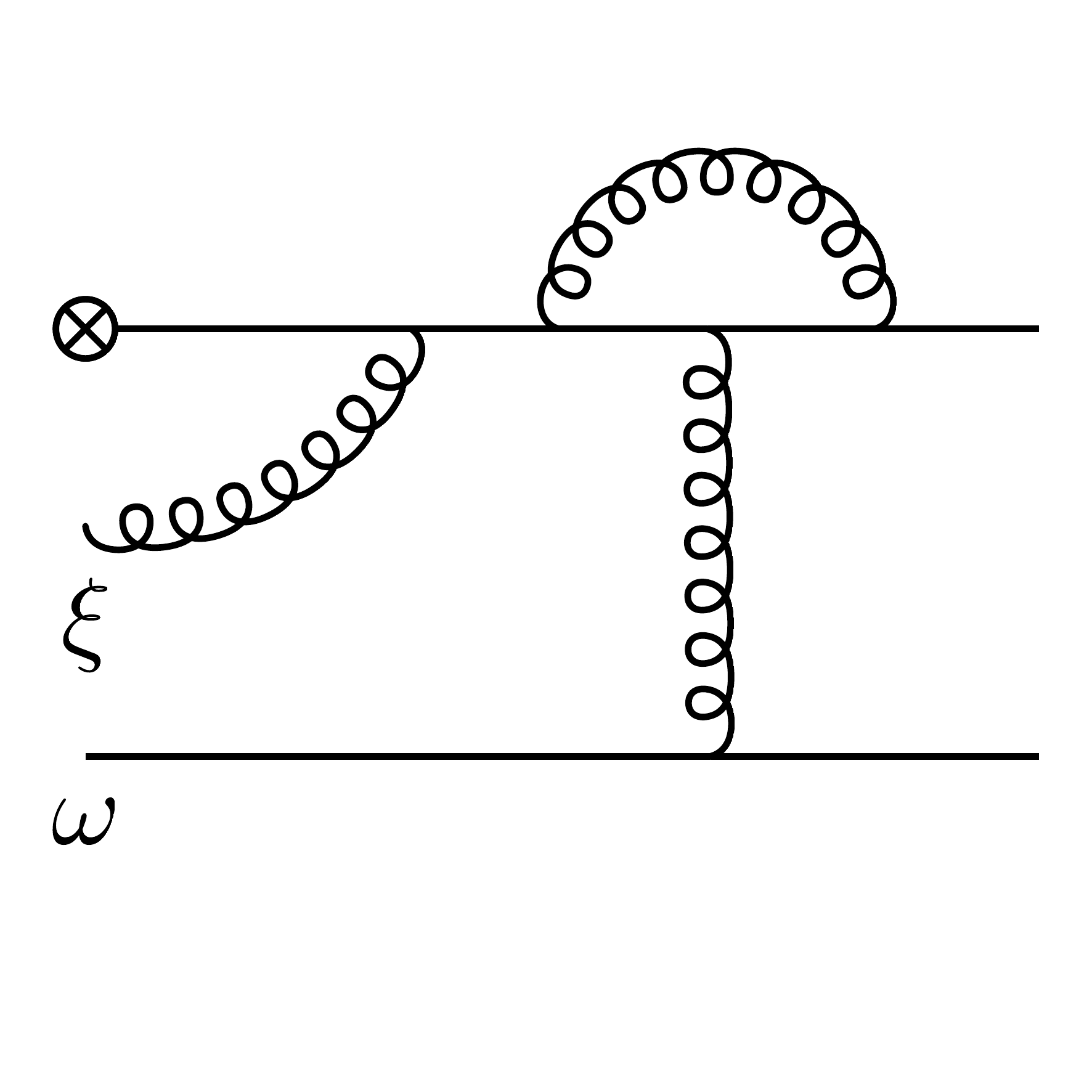}
    \includegraphics[trim = 20 100 20 100,clip,width = 0.24\textwidth]{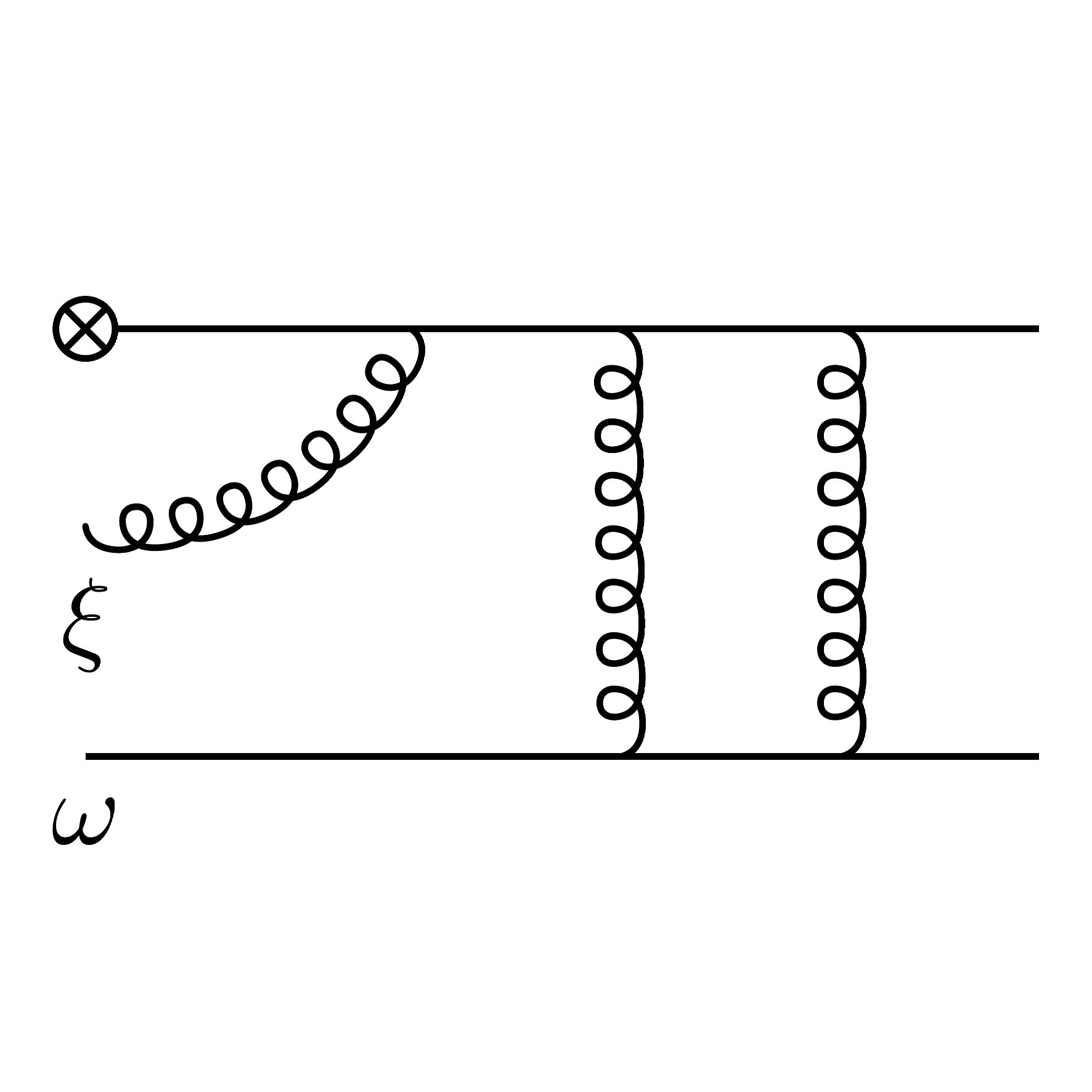}
    \includegraphics[trim = 20 100 20 100,clip,width = 0.24\textwidth]{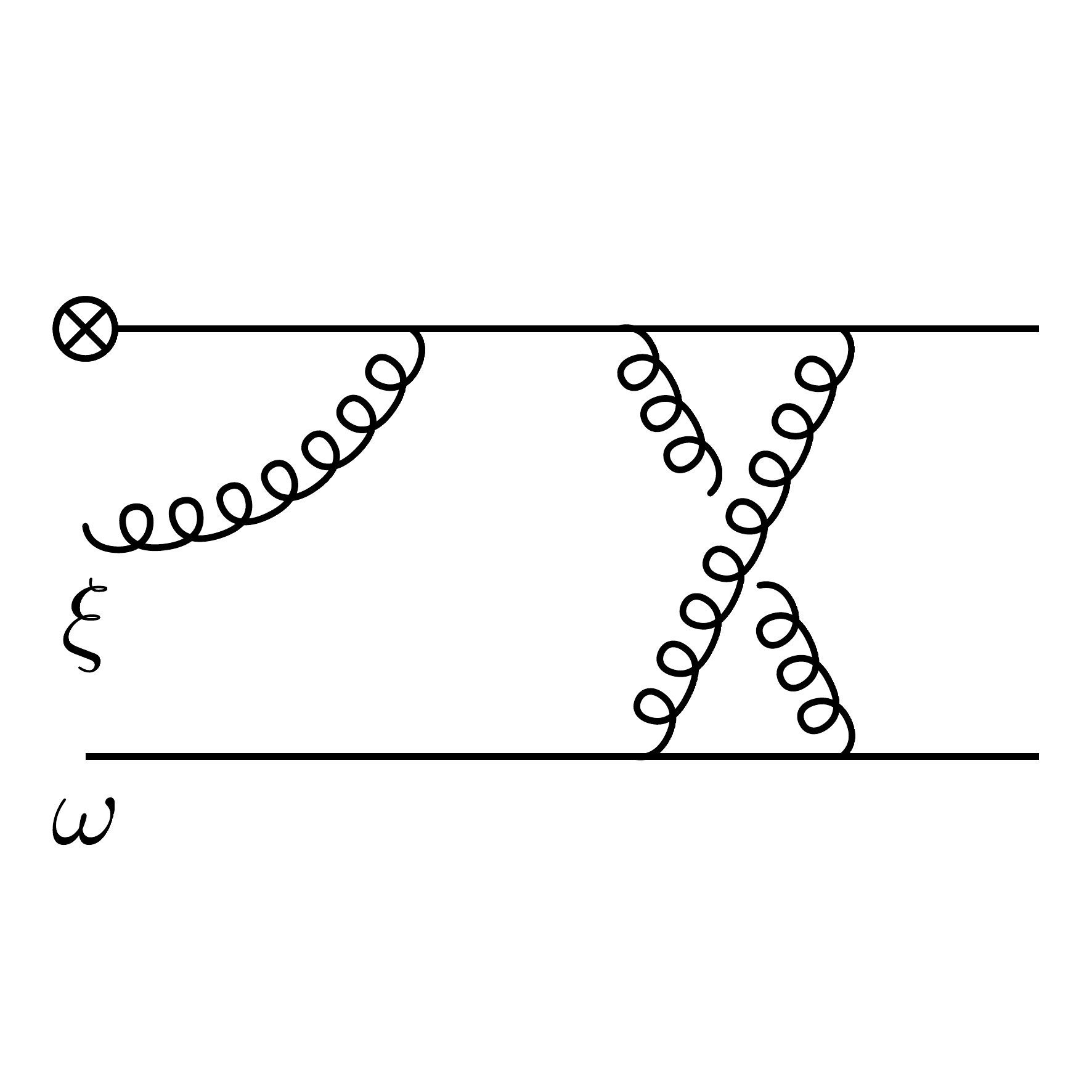}\\
    \includegraphics[trim = 20 100 20 100,clip,width = 0.24\textwidth]{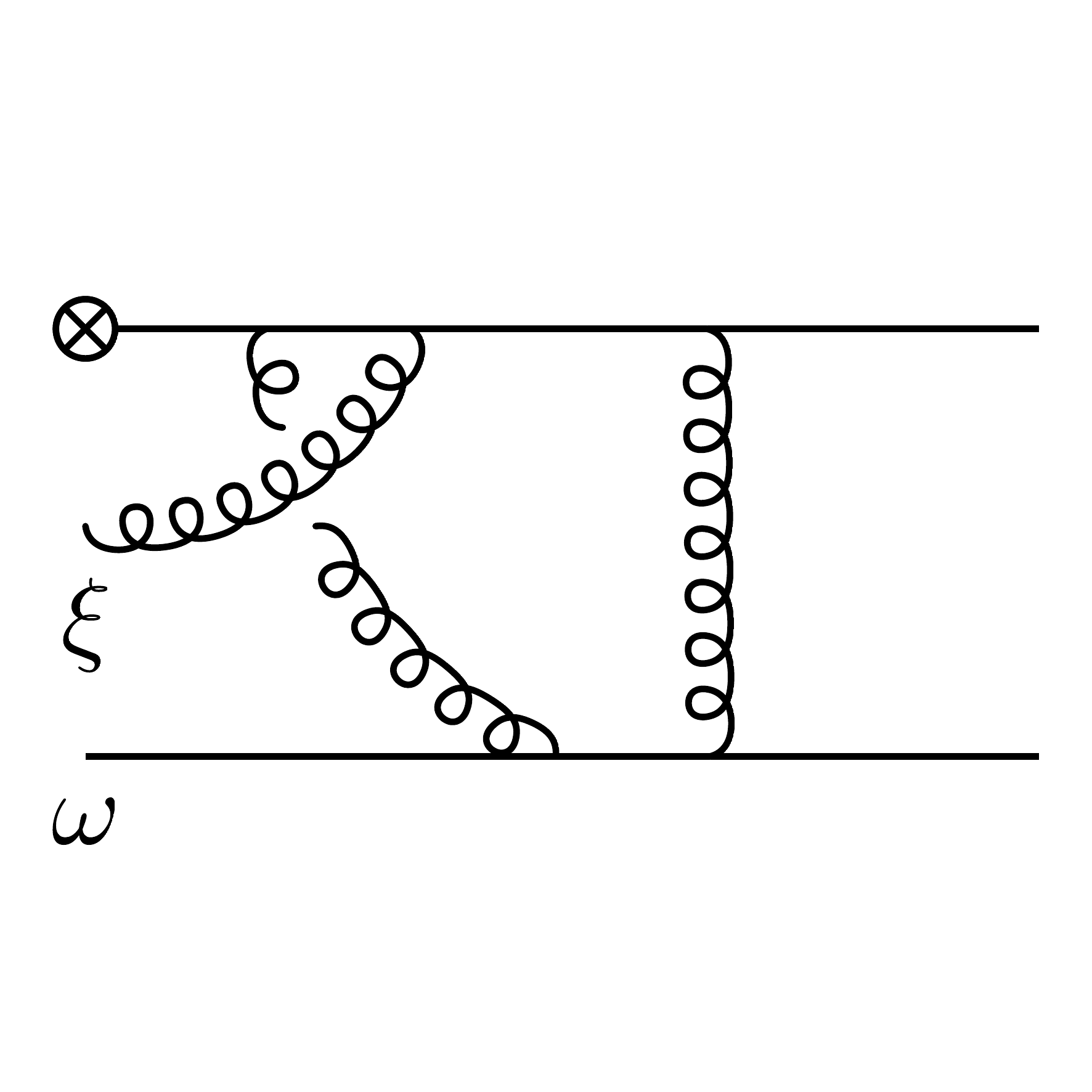}
    \includegraphics[trim = 20 100 20 100,clip,width = 0.24\textwidth]{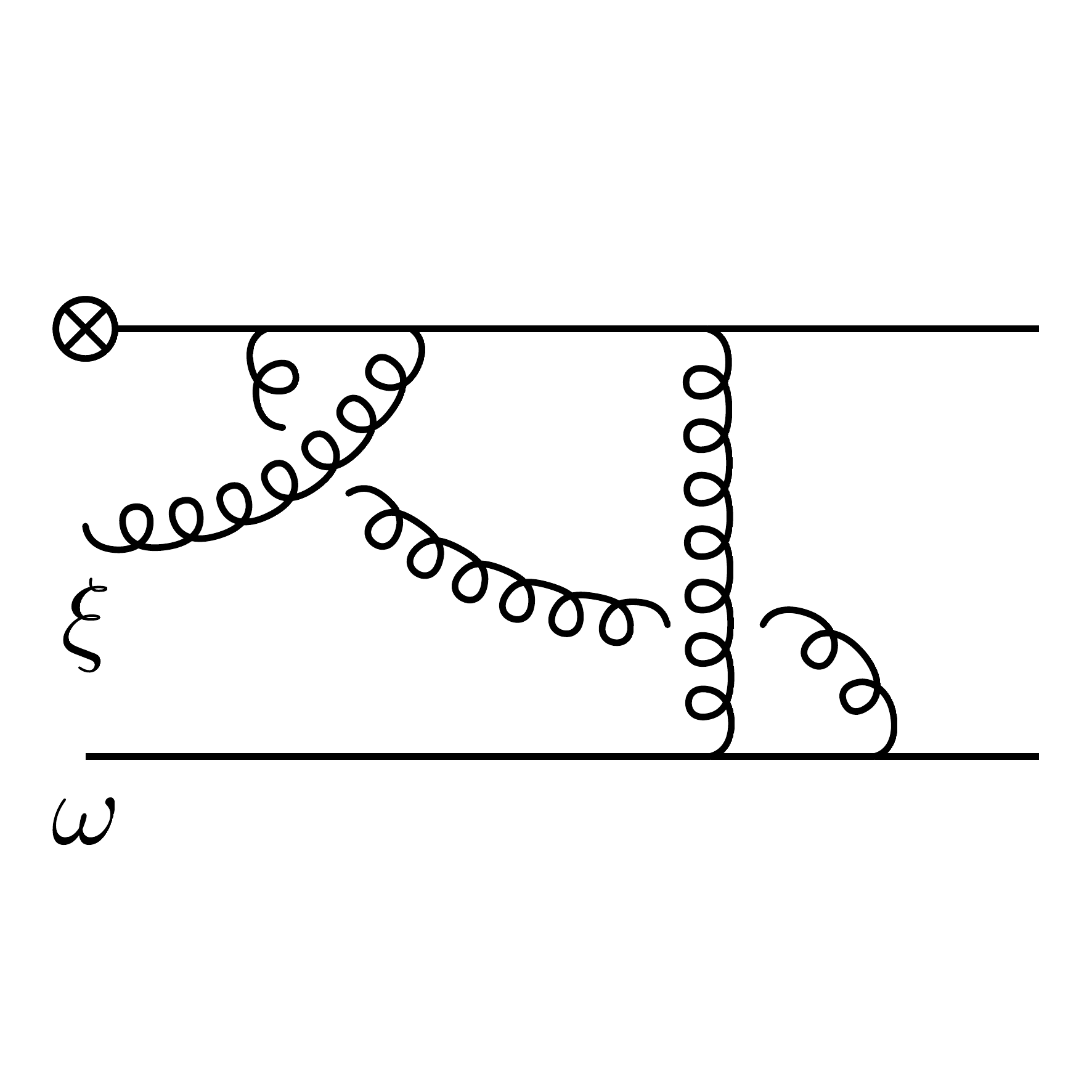}\\
	\caption{Sample one-loop corrections to the jet function $D_4(u,\omega,\xi)$. The first line shows diagrams that are only sensitive to the hard-collinear scale $(2\omega E_\eta)$, and the second line shows diagrams that are not as singular as $1/\omega$  for $\omega \to 0$. These contributions are irrelevant for the present discussion.
    }
    \label{fig:D41_diagrams}
\end{figure}

In this appendix, we get a better understanding of this observation from the field-theoretical point of view. For convenience, we will restrict our analysis to the Abelian limit. We begin by recalling that
\begin{align}
\label{eq:D40_smallomega}
    D_4^{(0)}(u,\omega,\xi)\big\vert_{C_A = 0} &= - \frac{C_F}{N_C} \, \frac{1}{4 E^2 \bar{u}^2 (\omega+\xi)^2} \frac{\bar{u} \xi  + \omega}{\omega} \nonumber \\
    &= - \frac{C_F}{N_C} \, \frac{1}{4 E^2 \bar{u} \omega \xi} + \mathcal{O}(\omega^0) \,, 
\end{align}
at tree-level, where the contribution in the second line arises entirely from the diagram in the middle of Fig.~\ref{fig:matching} in light-cone gauge. In the non-relativistic framework, $\phi_{3B}(\omega,\xi)$ does not vanish for $\omega \to 0$ (but for $\xi \to 0$) and the convolution integral needs to be performed with an additional rapidity regulator $\alpha$. More precisely~\cite{Bell:2008er},
\begin{align}
\label{eq:phi3BNLO}
    \phi_{3B}(\omega,\xi) &= 
    -\frac{\alpha_s C_F}{4\pi} \frac{\delta(\omega+\xi-m_2)}{m_2} \, \theta(m_2 - \xi) \, \xi^2 \left( \frac{1}{\varepsilon} + \ln \frac{\mu^2}{\xi^2} \right) + \mathcal{O}(\alpha_s^2) \,. 
\end{align}
Radiative corrections to $D_4$ must thus behave as $(2\xi E_\eta )^{-\varepsilon}/\omega$ for small $\omega$ in order to generate singularities in $\alpha$ from the convolution. We argue in the following that the contributions to 
$D_4^{(1)}(u,\omega,\xi) \propto (2\xi E_\eta)^{-\varepsilon}/\omega$ have only single poles in $\varepsilon$, and are thus irrelevant in the double-logarithmic approximation. 

At NLO, all diagrams that involve $(2\omega E_\eta)$ as the only hard-collinear scale in the loop can immediately be discarded, because the loop integral will generate a factor $\omega^{-\varepsilon}$ that results in $1/\varepsilon$ rather than $1/\alpha$ poles in the convolution. Examples are shown in the first line of Fig.~\ref{fig:D41_diagrams}. The situation is more involved for the diagrams shown in the second line, as the loop integrals involve both soft momenta $\omega$ and $\xi$. Since we are only interested in the singular piece $\sim 1/\omega$, one can use a method-of-region analysis adopting the scale hierarchy $\omega \ll \xi$ in this case. The low-energy momentum region will then produce a contribution from the scale $(2\omega E_\eta)$, which can be discarded for the same reasons as above. The ``large-energy'' momentum region of the integrals, on the other hand, is sensitive to the scale $(2\xi E_\eta)$ and needs to be taken into account. However, as the spectator quark directly attaches to the loop, it is immediately clear that these diagrams are power-suppressed in the limit $\omega \ll \xi$.

\begin{figure}[t]
    \centering
    \subfloat[]{
        \includegraphics[trim = 20 100 20 100,clip,width = 0.24\textwidth]{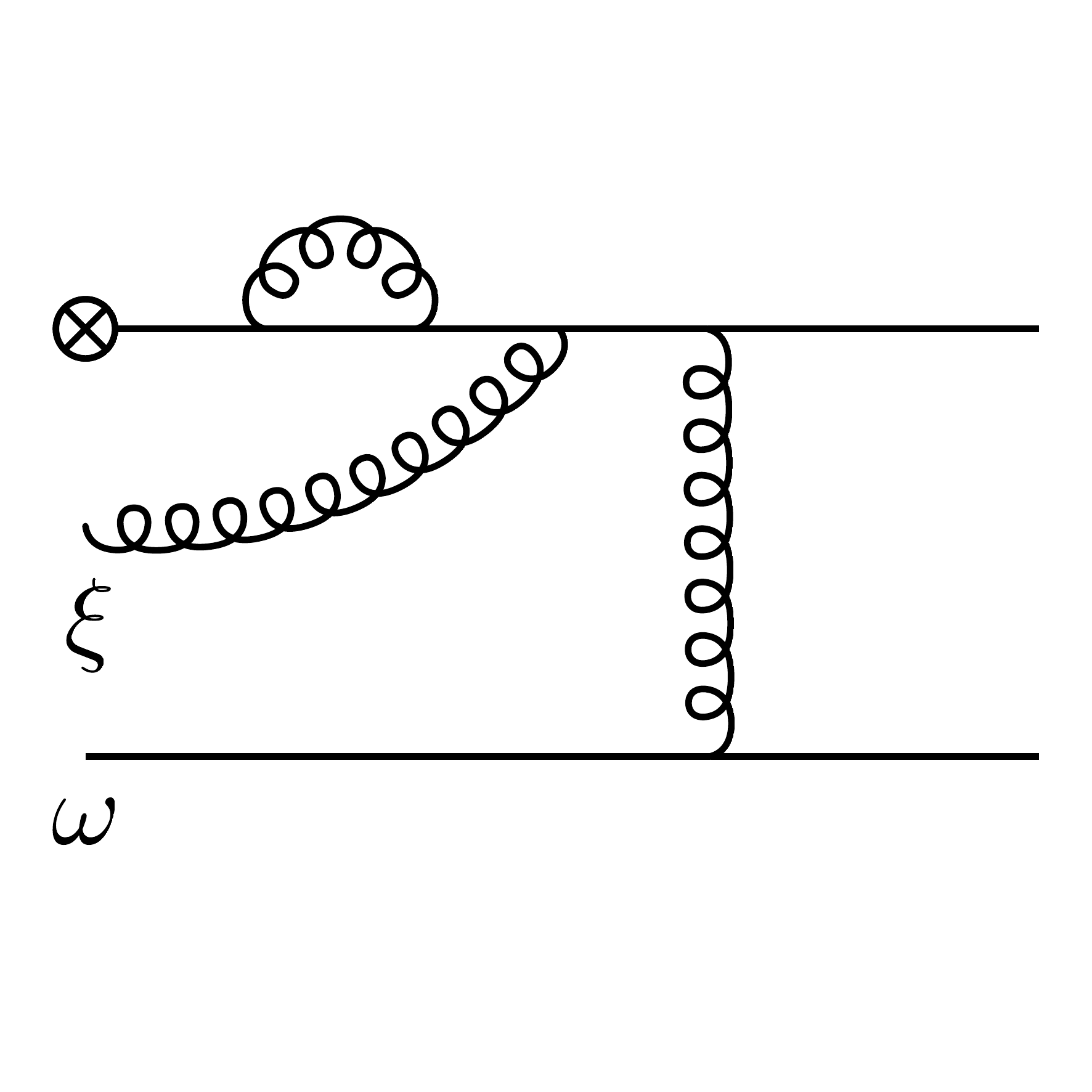}
    }
    \subfloat[]{
        \includegraphics[trim = 20 100 20 100,clip,width = 0.24\textwidth]{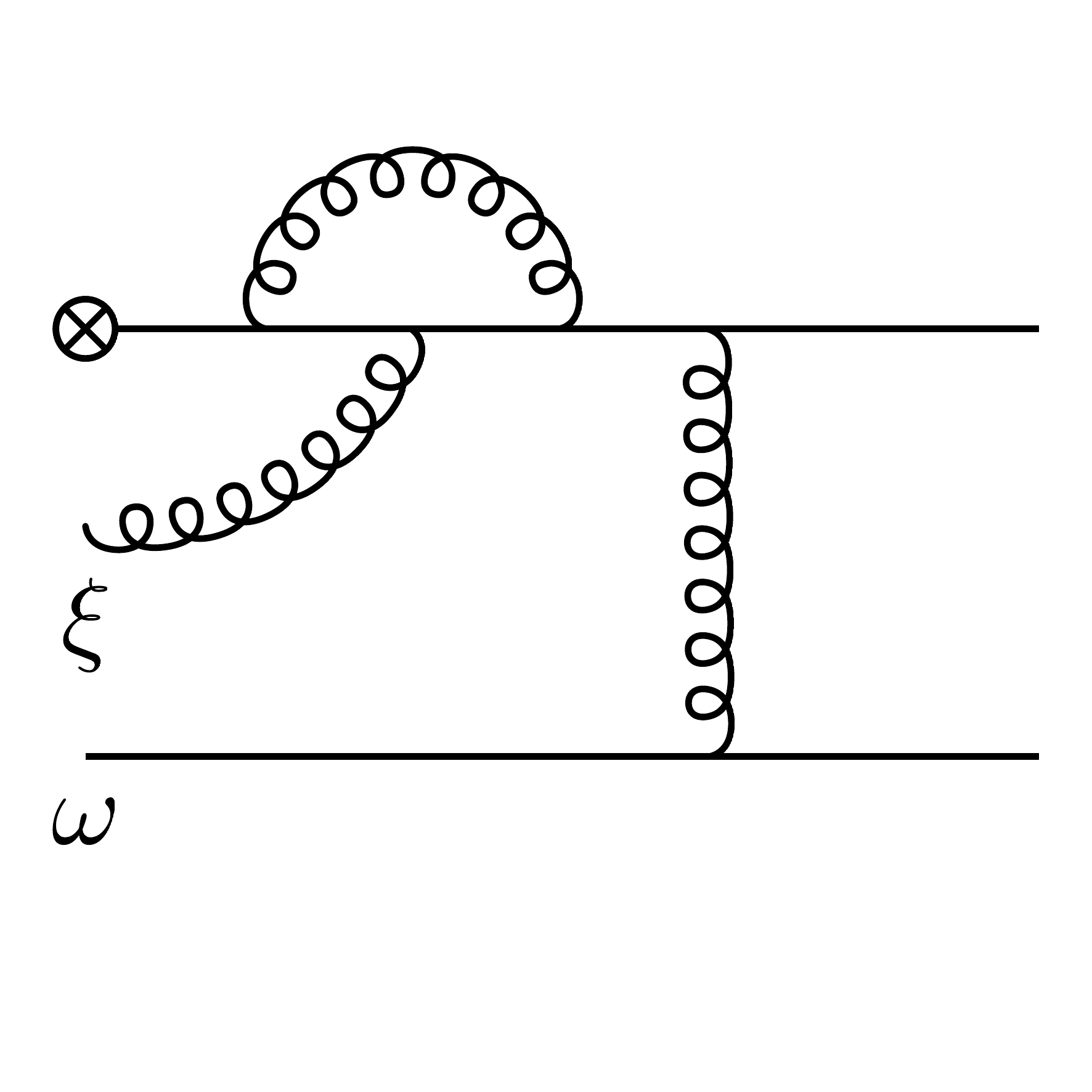}
    }
    \subfloat[]{
        \includegraphics[trim = 20 100 20 100,clip,width = 0.24\textwidth]{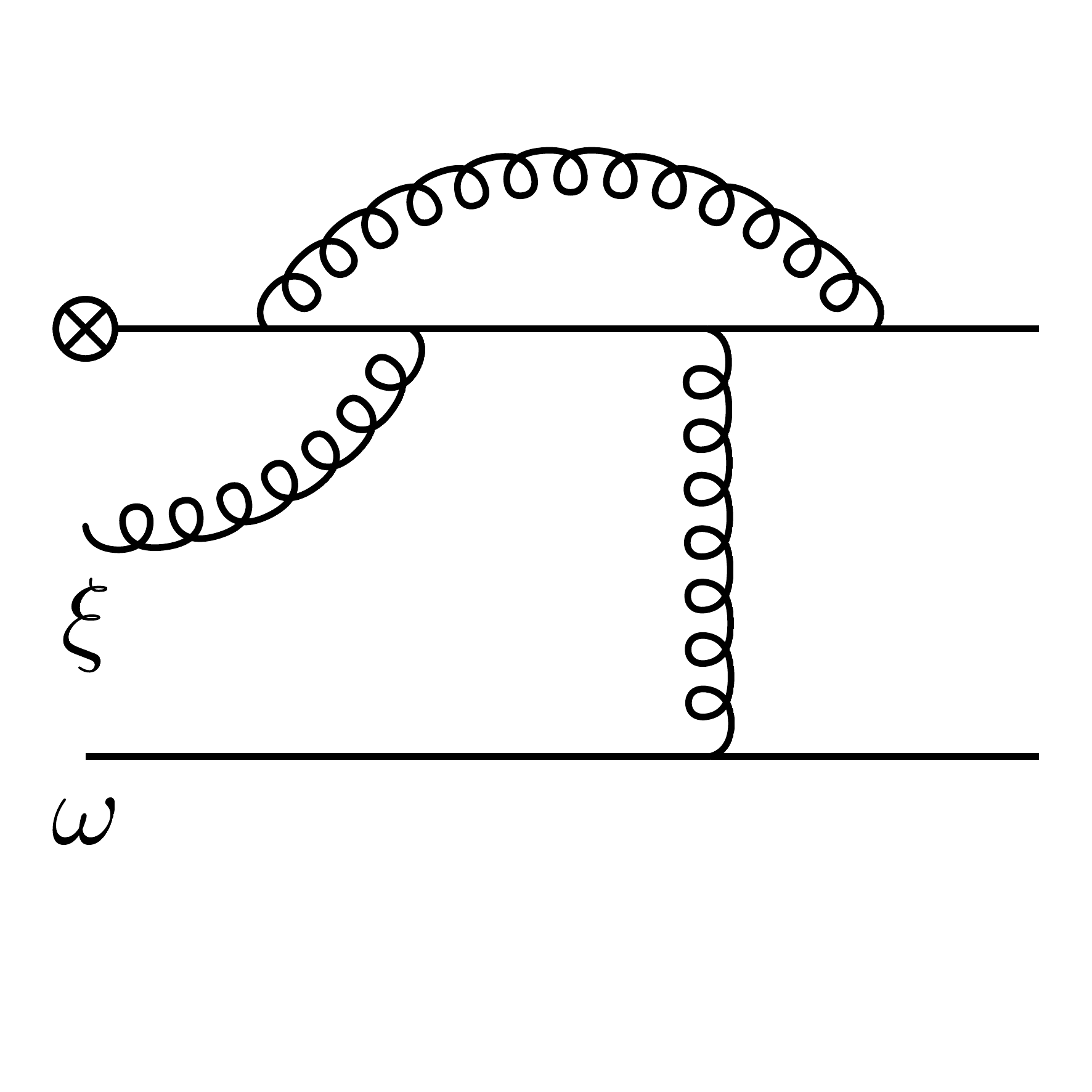}
    }
    \caption{Sample one-loop corrections to the jet function $D_4(u,\omega,\xi)$ that are relevant for the present discussion and need to be computed.}
    \label{fig:D41_diagrams_relevant}
\end{figure}

This leaves the three Abelian diagrams shown in Fig.~\ref{fig:D41_diagrams_relevant}. Here, the momentum $\omega$ can also be set to zero in the loop integral, but the additional factor $1/\omega$ is produced from the remaining propagators outside the loop. Furthermore, the loop integrals only depend on the scale $(2\xi E_\eta)$ and hence need to be computed. Their evaluation in collinear light-cone gauge shows that the diagram in the middle only produces a single pole in $\varepsilon$ and hence can be ignored in the double-logarithmic approximation. The remaining two diagrams both give rise to a double pole, which, however, cancels in the sum of the two contributions,
\begin{equation}
    D_4^{(1)}(u,\omega,\xi) \big\vert^{\text{Diag } a}_{C_A = 0} = - D_4^{(1)}(u,\omega,\xi) \big\vert^{\text{Diag } c}_{C_A = 0} \simeq - \frac{C_F}{N_C} \, \frac{1}{4 E^2 \bar{u} \omega \xi} \times \frac{4 C_F}{\varepsilon^2} \Big( \frac{\mu^2}{2\xi E_\eta} \Big)^\varepsilon + \ldots 
\end{equation}
where the dots refer to terms that are not of the $
(2\xi E_\eta)^{-\varepsilon}/\omega$ type in combination with double poles in $\varepsilon$. We hence confirmed by explicit calculation that all one-loop corrections to the jet function $D_4(u,\omega,\xi)$, which can potentially lead to a $1/\alpha$ divergence in the convolution with $\phi_{3B}^{(1)}(\omega,\xi)$, are single logarithmic. This explains the observation $(js)_{22} = (js)_{31} = 0$ in the Abelian limit. We note, however, that pole cancellation predicts the vanishing of these coefficients in full QCD.

Lastly, we comment that the above observations could be formalized at the operator level by refactorizing the relevant SCET-1 time-ordered products in the limit $\omega \to 0$ into two functions associated with the two scales $\omega \ll \xi$. A more formal analysis along these lines goes beyond the scope of this paper.

\section{Details of the \texorpdfstring{$j_{60}$}{}-fits}
\label{app:j60}

\begin{figure}[ht!]
				\centering
                \includegraphics[width=0.45\textwidth]{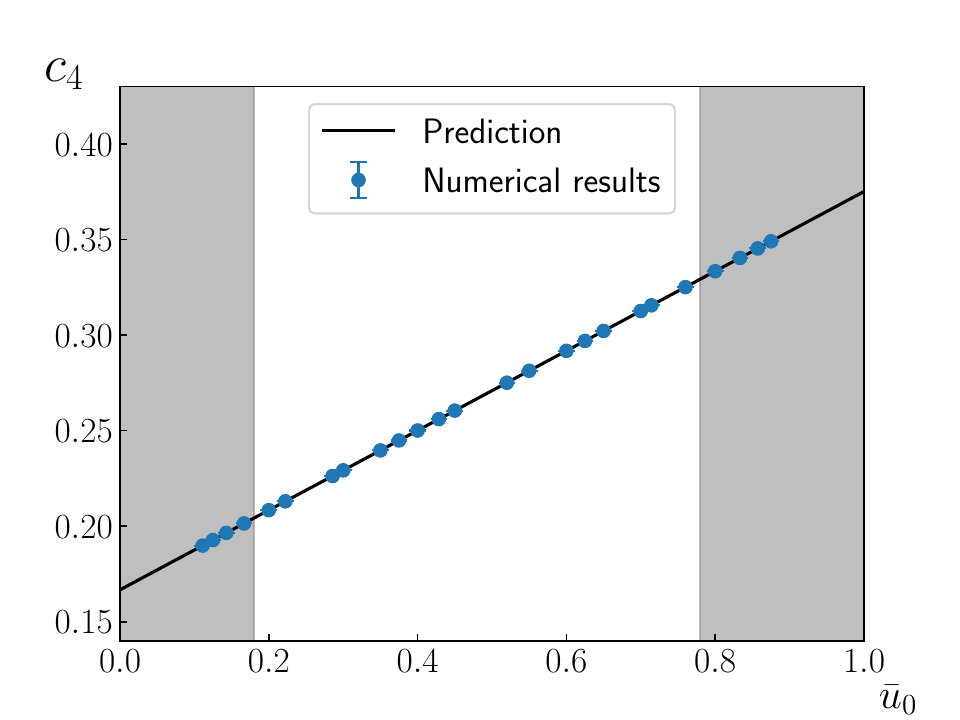}
                \includegraphics[width=0.45\textwidth]{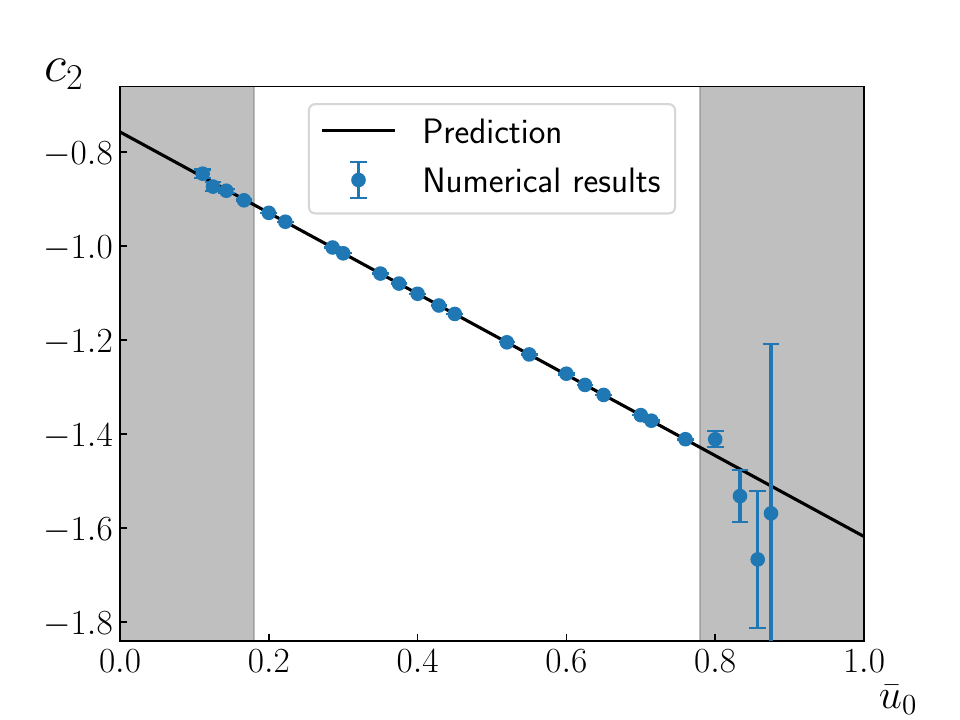}\\
                \includegraphics[width=0.45\textwidth]{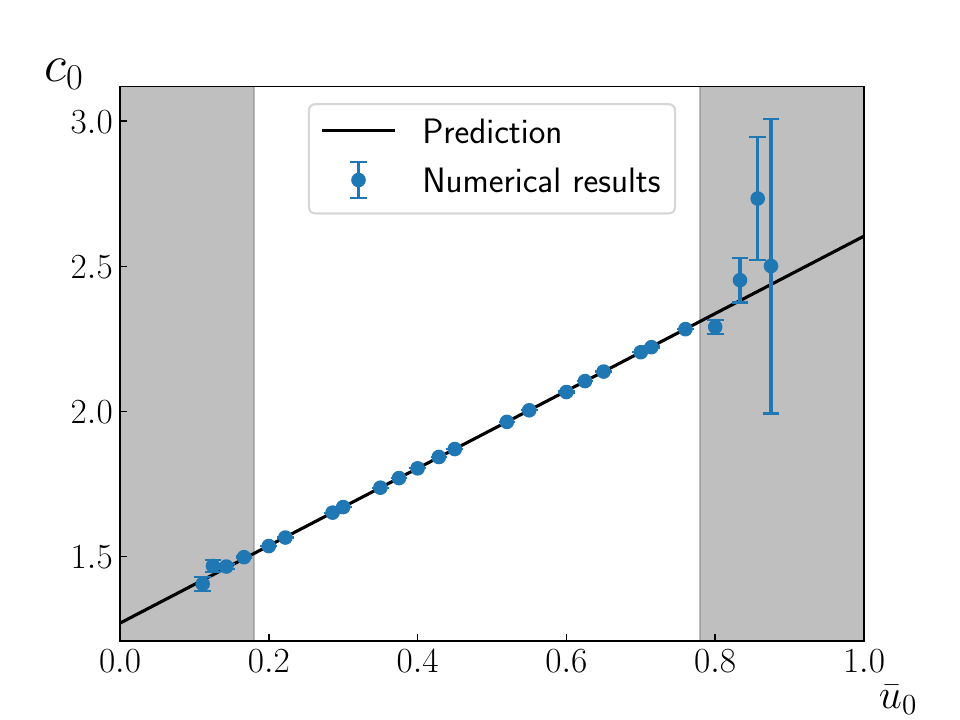}
                \includegraphics[width=0.45\textwidth]{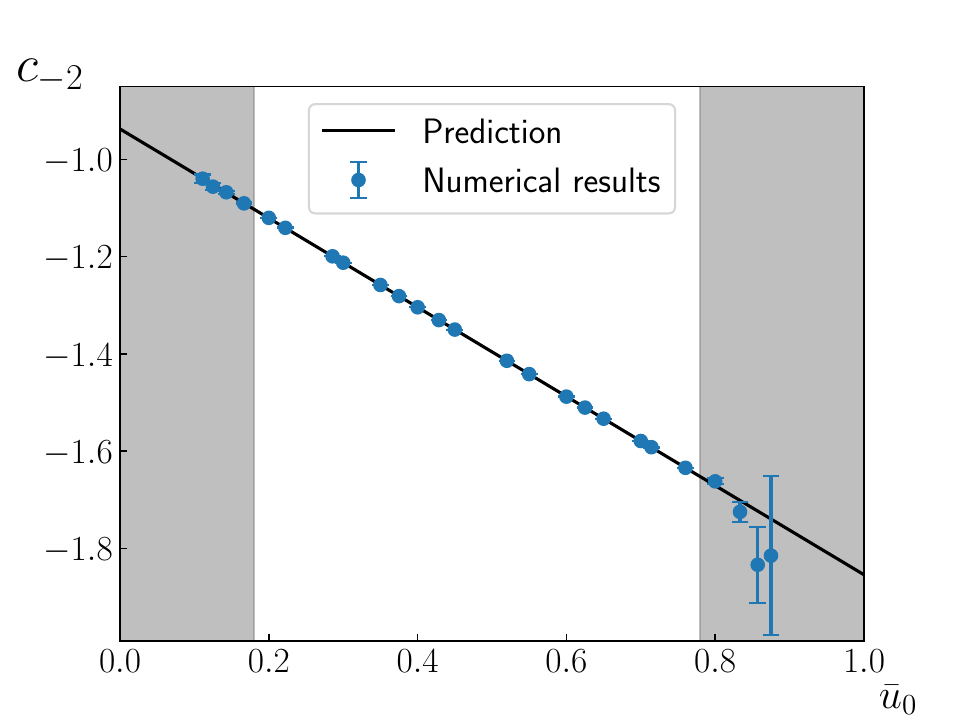}\\
                \includegraphics[width=0.45\textwidth]{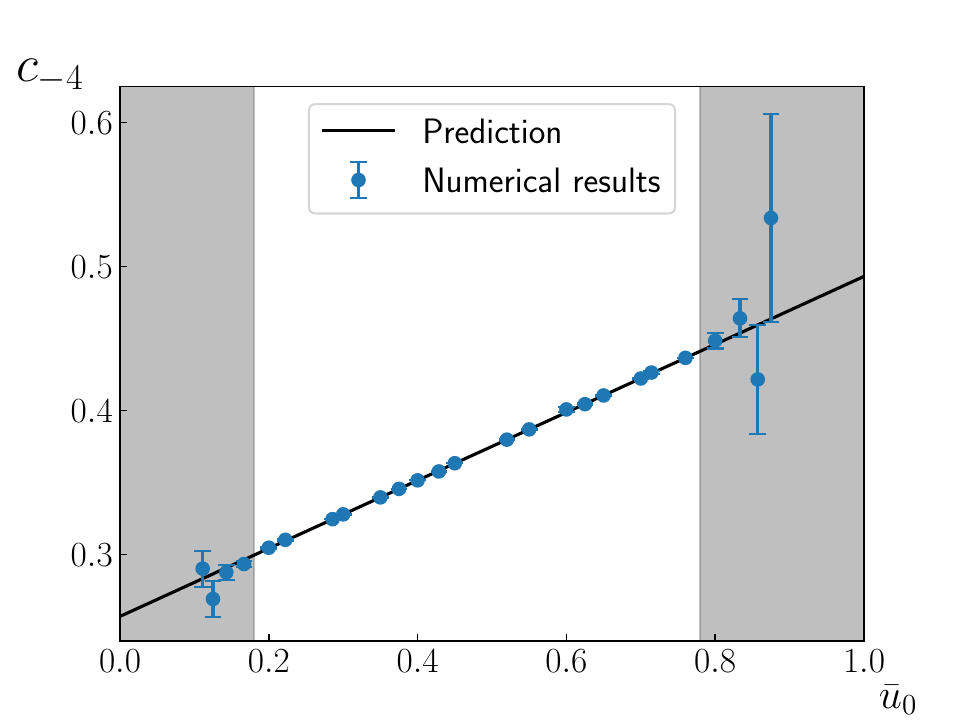}
	\caption{Results for the color coefficients defined in \eqref{eq:j60:fit:colour} as a function of the quark-mass ratio $\bar{u}_0$. The dots show the numerical results of our three-loop computation, and the lines display the expressions that are consistent with the integral equations, see \eqref{eq:j60:expectation}.
    }
	\label{fig:j60fit:colour} 
\end{figure}

In the main text we concluded that our numerical result for the hard-collinear coefficient $j_{60}$ in \eqref{eq:j60:fit} confirms the prediction of the integral equations at three-loop order. Here we report the fit results for the individual color structures. To do so, we write
\begin{align}
\label{eq:j60:fit:colour}
     \frac{j_{60}C_F\bar{u}_0^3}{4\xi_0} = c_4 \,N_c^4 + c_2 \, N_c^2 + c_0 + c_{-2} \,N_c^{-2} + c_{-4} \,N_c^{-4} \,.
\end{align}
Our results for the color coefficients $c_i$ are shown in Fig.~\ref{fig:j60fit:colour}, where the lines display the expression \eqref{eq:j60:expectation} that is predicted by the integral equations for $C_F=(N_c^2-1)/2N_c$ and $C_A=N_c$. The fit results clearly confirm this prediction for each of the color structures. 

\newpage
\bibliography{refs}
\end{document}